\documentclass[lettersize,journal]{IEEEtran}
\usepackage{amsmath,amsfonts,amssymb,amsthm}
\usepackage{mathtools}
\usepackage{algorithmic}
\usepackage{algorithm}
\usepackage{array}
\usepackage{booktabs}
\usepackage{multirow}
\usepackage[caption=false,font=normalsize,labelfont=sf,textfont=sf]{subfig}
\usepackage{textcomp}
\usepackage{stfloats}
\usepackage{url}
\usepackage{verbatim}
\usepackage{graphicx}
\graphicspath{{figures/}}
\usepackage{hyperref}
\usepackage{flushend}
\usepackage[capitalize,noabbrev]{cleveref}

\def\BibTeX{{\rm B\kern-.05em{\sc i\kern-.025em b}\kern-.08em
    T\kern-.1667em\lower.7ex\hbox{E}\kern-.125emX}}
\usepackage{balance}
\usepackage[dvipsnames,table]{xcolor}
\usepackage{colortbl}
\newcommand{\gray}[1]{\textcolor{gray}{#1}}

\providecommand{\Xhline}[1]{\noalign{\hrule height #1\relax}}
\newcommand{\best}[1]{\textcolor{red!85!black}{\textbf{#1}}}
\newcommand{\second}[1]{\textcolor{blue!75!black}{\textbf{#1}}}
\newcommand{\std}[1]{$_{\pm\text{\scriptsize #1}}$}

\newcommand{\venue}[1]{\,\textcolor{gray!60!black}{\scriptsize{[#1]}}}
\newcommand{\method}{TiLP}
\begin{document}
\title{Application-Aware Twin-in-the-Loop Planning for Federated Split Learning over Wireless Edge Networks}
\author{Zihao~Ding,
        Beining~Wu,~\IEEEmembership{Member,~IEEE},
        Jun~Huang,~\IEEEmembership{Senior Member,~IEEE}, and
        Shiwen~Mao,~\IEEEmembership{Fellow,~IEEE}%
\IEEEcompsocitemizethanks{
\IEEEcompsocthanksitem Zihao Ding, Beining Wu, and Jun Huang are with the Department of Electrical Engineering and Computer Science, South Dakota State University, Brookings, SD 57007, USA.
E-mails: \{Zihao.Ding, Wu.Beining\}@jacks.sdstate.edu, Jun.Huang@sdstate.edu.
\IEEEcompsocthanksitem Shiwen Mao is with the Department of Electrical and Computer Engineering, Auburn University, Auburn, AL 36849, USA.
E-mail: smao@auburn.edu.
}}

\markboth{IEEE Journal on Selected Areas in Communications,~Vol.XX, No.~X, XXXX~2026}
{IEEE Journal on Selected Areas in Communications}

\maketitle

\begin{abstract}
We investigate task-success-oriented resource allocation for federated split learning (FSL) at the wireless edge. In this setting, the server must jointly determine bandwidth, transmit power, split-layer placement, compression level, and terminal participation under per-round deadline, memory, and spectrum constraints. These coupled decisions affect wireless transmission, model training, and task execution, which evolve at different time scales and cannot be efficiently evaluated through repeated real-world trials. To address this challenge, we propose \method{}, a twin-in-the-loop planner that evaluates candidate decisions through a cross-domain digital twin before execution. The twin integrates network, training, and task sub-twins, with each sub-twin calibrated at the time scale of the process it models. Based on this twin, \method{} performs receding-horizon cross-entropy method planning with actor-critic guidance to search over mixed continuous-discrete decisions. Experiments on LIBERO robotic manipulation tasks over an Sionna RT-simulated wireless network show that \method{} improves task success by 9.5 percentage points over the strongest single-axis baseline, while satisfying the per-round deadline and energy budget.

\end{abstract}


\section{Introduction}
\label{sec:introduction}

\IEEEPARstart{F}{oundation} models are increasingly being adapted at the wireless edge, where terminals often lack the computation and memory needed to fine-tune an entire model locally. Federated split learning (FSL) provides a practical architecture for this setting: each terminal executes only the front layers of the model, while a co-located edge server executes the remaining layers~\cite{Singh2017PMLR, Praneeth2018arXiv, Thapa2022AAAI, Wu2026MNET}. This architecture reduces the local burden on terminals and avoids direct transmission of raw data. However, it shifts the system design challenge from model placement alone to per-round resource allocation. In each training round, the edge server must decide which terminals participate, how much bandwidth and transmit power they receive, where the model is split, and how much the exchanged activations and gradients are compressed ~\cite{Wu2026COMST,DingICNC2025,Wu2025ToN}.

These decisions are tightly coupled. A deeper split layer may reduce terminal-side computation but increase the activation payload. Stronger compression can reduce communication delay but may degrade the model update. Scheduling more terminals can improve data diversity but may also increase straggling under limited spectrum. Therefore, resource allocation in wireless FSL cannot be treated as a collection of independent decisions; it must be optimized jointly under communication, computation, memory, and deadline constraints.

The objective of this joint allocation becomes more demanding when the fine-tuned model is deployed for a terminal-side control task, such as robotic manipulation. Existing wireless FSL methods typically optimize training loss, gradient divergence, latency, or energy. These metrics are useful, but they are not reflect the performance of the deployed policy. A lower training loss does not necessarily mean that a robot will grasp an object, insert a tool, or complete a long-horizon manipulation task successfully~\cite{Wu2026TNSE}. Thus, the server should not only make training faster or communication efficient, it should allocate resource that improve task success. In this sense, the task-success objective fundamentally changes wireless FSL from a training-efficiency problem into a cross-domain planning problem.



However, the challenge is that, task-success-oriented control couples three processes that are usually treated separately. The first process is wireless communication. Bandwidth and power decisions determine the transmission rate, communication delay, and energy consumption in each round. The second process is split learning. Split-layer and compression decisions affect activation size, memory usage, gradient quality, and model updates. The third process is task execution. The trained model must be evaluated through task rollouts, which are expensive and cannot be performed for every candidate allocation. These three processes operate at different time scales and interact through a high-dimensional mixed action space. As a result, three questions must be answered.

First, \emph{how should per-round FSL resource allocation be optimized when training loss is not a reliable proxy for deployment performance?} Existing formulations commonly use training loss, gradient divergence, latency, or energy as the main objective. Although these metrics are important, they do not reflect whether the learned policy succeeds in deployment.

Second, \emph{how can the task-level effect of a candidate allocation be estimated without testing every candidate on the physical system?} Direct evaluation is impractical because each candidate consumes resources, changes the model, and may require expensive task rollouts before its value is known.

Third, \emph{how can the controller plan over a mixed continuous-discrete action space that spans wireless control, split-learning dynamics, and task evaluation?} The action includes continuous variables, such as bandwidth, power, and compression, as well as discrete variables, such as split-layer selection and terminal scheduling. A planner must reason over these variables jointly while enforcing system constraints.

Prior work addresses only parts of this problem. Wireless FSL resource-allocation methods optimize selected variables, such as split layers, compression ratios, bandwidth, power, or client schedules, but typically use training loss or gradient divergence as the objective~\cite{Xu2024TWC, Lin2025TMC_a, Hou2025JSAC, Wang2025TWC, Wang2026TWC,Wu2023MPE,Wu2025RACS,Ding2025IPCCC,Pudasaini2026HPSR}. Digital twins for wireless and edge systems provide useful prediction tools, but existing twins usually focus on a single dimension, such as radio propagation, training dynamics, or task simulation~\cite{Khan2025TWC, Okegbile2025TMC, Zhou2025TMC, Ho2025JSAC,Wu2026ICDCS,Ding2026ICDCS}. They do not provide a unified toolkit that predicts the joint effect of an allocation on communication feasibility, learning progress, and task success. Model-based reinforcement learning (MBRL) offers a natural planning framework~\cite{Hafner2019ICML, Hafner2025NATURE, Chua2018NeurIPS, Janner2019NeurIPS,Huang2025TMC,Wu2026ARXIV}, but standard models are usually designed for a single dynamical process and predominantly continuous action spaces. The underlying assumptions do not hold in wireless FSL, where channel variation, model training, and task evaluation evolve at different time scales and the control variables are intrinsically mixed.

To address these limitations, we propose \method{}, a Twin-in-the-Loop Planner for task-success-oriented FSL at the wireless edge. TiLP formulates the per-round control problem as a Markov decision process whose reward is driven by task-success improvement, rather than training loss alone. To evaluate candidate actions safely and efficiently, TiLP builds a cross-domain digital twin composed of three coupled sub-twins: a network sub-twin for wireless rate, latency, and energy prediction; a training sub-twin for split-learning progress prediction; and a task sub-twin for estimating task success. Each sub-twin is calibrated at the time scale of the process it models: per round for wireless dynamics, per aggregation interval for learning dynamics, and per task-evaluation interval for task performance.

On top of this cross-domain twin, TiLP performs receding-horizon planning using the cross-entropy method (CEM). The planner searches directly over the full mixed action space, including bandwidth, power, compression, split-layer selection, and terminal scheduling. A learned actor initializes the CEM sampling distribution, while a soft critic provides a terminal value estimate beyond the planning horizon. When the per-round computation budget is limited, the actor can be used alone as a lightweight deployment policy.

The main contributions of this paper are summarized as follows.

\begin{itemize}
    \item We formulate task-success-oriented resource allocation for foundation-model FSL over wireless edge networks. The formulation jointly controls bandwidth, transmit power, split-layer location, compression level, and terminal scheduling under deadline, memory, spectrum, and energy constraints.

    \item We develop a cross-domain digital twin that integrates network, training, and task sub-twins into a unified planning surrogate. The three sub-twins are calibrated at different time scales to match wireless dynamics, split-learning updates, and task-level evaluations.

    \item We propose TiLP, a receding-horizon planner that combines CEM with actor-critic guidance to search over mixed continuous-discrete FSL actions. This design enables online planning with twin-based evaluation and provides a lightweight actor-only fallback for deployment under limited computation.

    \item We evaluate TiLP on LIBERO manipulation tasks with FSL deployed over a Sionna RT-simulated wireless network, where it improves task-success rate by 9.5 percentage points over the strongest single-axis baseline, while staying within the per-round deadline and energy budget.
\end{itemize}

The rest of this paper is organized as follows. Section~\ref{sec:related_work} reviews related work. Section~\ref{sec:sys} presents the system model and problem formulation. Section~\ref{sec:method} introduces TiLP. Section~\ref{sec:exp} reports the experimental results. Section~\ref{sec:conclusion} concludes the paper.

\section{Related Work}
\label{sec:related_work}

\subsection{Federated Split Learning over Wireless Networks}

Federated split learning (FSL) enables edge terminals to train large models without storing the entire model locally~\cite{Singh2017PMLR, Praneeth2018arXiv, Thapa2022AAAI,Wu2026MNET,Fang2025ARXIV,Fang2026GLOBECOM}. The client runs the front part of the model, the server runs the remaining part, and activations and gradients are exchanged at the split layer. This design reduces on-device computation, but it introduces a coupled control problem: the system must decide the split layer, compression level, terminal schedule, and wireless resource allocation.

Existing FSL studies address different parts of this problem. Some works reduce activation and gradient payloads through adaptive quantization, low-rank approximation, and compression~\cite{Mu2025TMC, Ao2026TWC, Qiao2024JSAC,Wu2023ACCESS,Pan2023SCIS}. Others optimize model partitioning through hierarchical or adaptive split placement~\cite{Lin2025TMC_a, Lin2025TMC_b}. More recent works incorporate wireless-side decisions, including bandwidth and power control~\cite{Xu2024TWC, Ao2025TCE, Liang2026TWC}, bandit-based partitioning~\cite{You2026TMC}, client selection~\cite{Xie2025TWC, Wang2024TON}, and latency minimization~\cite{Wen2025TNSE}. These efforts have advanced wireless FSL, but most of them optimize one control dimension while simplifying the others.

Recent studies on edge fine-tuning of foundation models follow a similar pattern~\cite{Hou2025JSAC, Wang2025TWC, Wang2026TWC,Xing2026ACR}. Their objectives mainly focus on training loss, gradient divergence, communication cost, or energy consumption. These metrics are useful, but they do not directly measure whether a deployed policy succeeds. For manipulation and embodied-AI tasks, lower training loss does not necessarily imply higher task success. This paper addresses this gap by jointly considering wireless control, split-learning dynamics, and task-level success.

\subsection{Digital Twins for Wireless and Edge Systems}

Digital twins allow a controller to evaluate candidate decisions before applying them to the physical system. Existing twins for wireless edge systems usually focus on one layer. Radio-layer twins model channels, propagation, rate, or SNR~\cite{Khan2025TWC, Cao2025TWC, Zhang2024JSAC,Wu2026ARXIV1,Fang2025TON,Fang2025JSAC}. Training-layer twins predict loss, gradients, or convergence behavior in distributed learning~\cite{Okegbile2025TMC, Jin2026TNSE, Chen2025TMC}. Task-layer twins evaluate learned policies in virtual environments and report task-level outcomes~\cite{Zhou2025TMC, Ho2025JSAC, Chen2024TMC}.

These twins are useful but incomplete for wireless FSL. A radio twin cannot predict training improvement. A training twin cannot determine task success. A task twin usually ignores the wireless and split-learning decisions that produced the model. Some recent methods use twins mainly as data generators for model-free policy training~\cite{Chen2025TMC, Chen2024TMC}, but they do not provide an explicit cross-domain planning surrogate. In contrast, this paper uses a digital twin that integrates network, training, and task sub-twins, with each sub-twin calibrated at its own time scale.

\subsection{Model-Based Reinforcement Learning with World Models}

Model-based reinforcement learning (MBRL) uses a learned model to simulate future outcomes before acting. One line of work learns latent world models for planning and policy learning, such as PlaNet~\cite{Hafner2019ICML} and Dreamer~v3~\cite{Hafner2025NATURE}. Another line uses the model as an explicit rollout engine, as in PETS~\cite{Chua2018NeurIPS} and MBPO~\cite{Janner2019NeurIPS}, often combined with CEM~\cite{Rubinstein1999} or actor-critic learning~\cite{Haarnoja2018ICML}. Recent studies also use twins or generative models for policy improvement~\cite{Tong2025TMC, Liu2024TMC}.

However, standard MBRL is not designed for the wireless FSL setting. First, most world models describe a single homogeneous process, while wireless FSL couples channel dynamics, split-learning updates, and task evaluation at different time scales. Second, many planners mainly target continuous action spaces, whereas our action includes continuous bandwidth, power, and compression variables, categorical split-layer decisions, and binary scheduling decisions. Our design addresses these challenges by placing a cross-domain digital twin inside a receding-horizon planner and searching over the full mixed action space.

\section{System Model and Problem Formulation}
\label{sec:sys}

\subsection{System Overview}
\label{sec:sys-overview}
\label{sec:sys-taskdt}

We consider an edge AI system that consists of a base station (BS), a co-located edge server, and a set of wireless terminals. Let $\mathcal{N}\triangleq\{1,\ldots,N\}$
denote the set of terminals. These terminals collaboratively fine-tune a shared foundation model $\Phi$ using their own private data. Since the data remain at the terminals, the training process follows federated split learning (FSL).

In FSL, the model $\Phi$ is divided into two parts. The front part is stored and trained on each terminal, while the remaining part is stored and trained at the edge server. During training, a terminal runs the front part of the model and sends the intermediate output, called the split-layer activation, to the server. The server then runs the remaining layers, computes the loss, and sends the gradient at the split layer back to the terminal.

The training process has $T$ rounds. Let $\mathcal{T}\triangleq\{1,\ldots,T\}$ denote the set of training rounds. Each round must be completed before a deadline. Terminal $n$ has a maximum transmit power $P^{\max}$ and an on-device memory budget $M_n^{\max}$. Since the exchanged activations and gradients can be large, especially when the split layer is deep, we allow lossy compression to reduce the communication load. Let $q^{\max}$ denote the maximum allowed compression level.

In each round $t$, the BS makes five decisions for each terminal $n$:
\begin{equation}
    \mathbf{a}_n[t]
    \triangleq
    \bigl(b_n[t],p_n[t],\ell_n[t],q_n[t],x_n[t]\bigr),
    \label{eq:action-terminal}
\end{equation}
where $b_n[t]$ is the bandwidth allocated to terminal $n$, $p_n[t]$ is its transmit power, $\ell_n[t]$ is the split-layer index, $q_n[t]$ is the compression level, and $x_n[t]\in\{0,1\}$ is the scheduling indicator. If $x_n[t]=1$, terminal $n$ is scheduled in round $t$. If $x_n[t]=0$, it is not scheduled.

The system-level action in round $t$ is
\begin{equation}
    \mathbf{a}[t]
    \triangleq
    \{\mathbf{a}_n[t]\}_{n\in\mathcal{N}}.
    \label{eq:action-system}
\end{equation}
The set of scheduled terminals is
\begin{equation}
    \mathcal{S}[t]
    \triangleq
    \{n\in\mathcal{N}:x_n[t]=1\}.
    \label{eq:scheduled-set}
\end{equation}

If terminal $n$ is scheduled and chooses split layer $\ell_n[t]$, then the terminal executes the layers from the input layer up to layer $\ell_n[t]$. The server executes the remaining layers. The activation produced at layer $\ell_n[t]$ is compressed and sent to the server. The compression level $q_n[t]$ represents the fraction of the payload that is discarded. Therefore, only the fraction $1-q_n[t]$ of the activation and gradient payload is transmitted. A larger $q_n[t]$ reduces communication time, but it also removes more information from the exchanged tensors.

Directly testing an action $\mathbf{a}[t]$ on the physical system can be risky. A bad action may cause a terminal to miss the round deadline, produce a poor model update, or reduce the task performance before the system can correct it. To avoid unsafe trial-and-error on the physical system, we construct a cross-domain digital twin (DT):
\begin{equation}
    \widehat{\Omega}[t]
    =
    \bigl(
    \widehat{\Omega}^{\mathrm{net}}[t],
    \widehat{\Omega}^{\mathrm{tr}}[t],
    \widehat{\Omega}^{\mathrm{task}}[t]
    \bigr).
    \label{eq:dt-overview}
\end{equation}
The DT contains three coupled sub-twins. The network sub-twin $\widehat{\Omega}^{\mathrm{net}}[t]$ models the wireless transmission environment. The training sub-twin $\widehat{\Omega}^{\mathrm{tr}}[t]$ predicts how much training improvement one round can produce. The task sub-twin $\widehat{\Omega}^{\mathrm{task}}[t]$ estimates the task success rate of the current model.

\begin{figure}[t]
    \centering
    \includegraphics[width=\linewidth]{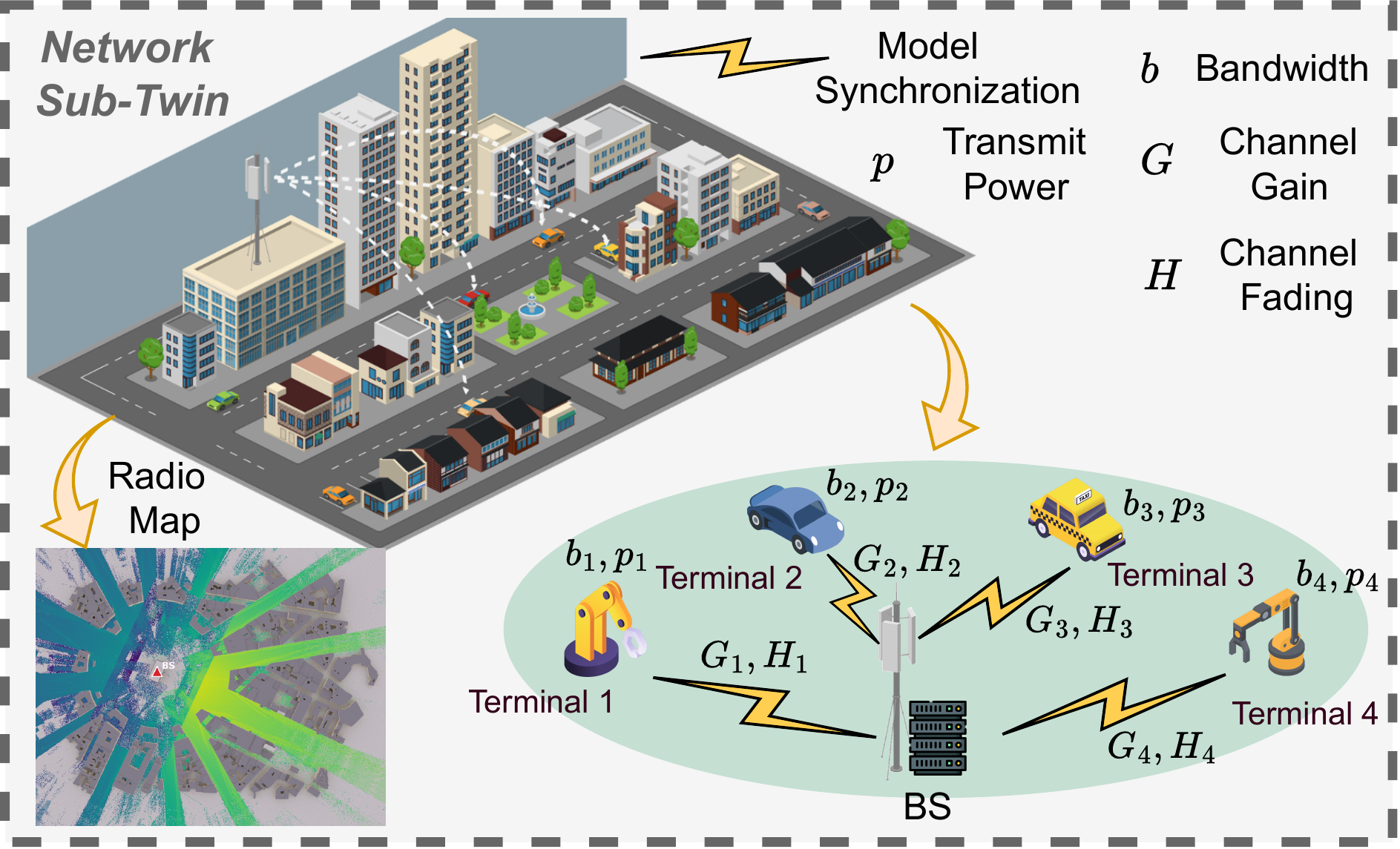}
    \caption{Network sub-twin illustration.}
    \label{fig:network-twin}
\end{figure}

\subsection{Network Sub-Twin}
\label{sec:sys-wireless}

The purpose of the network sub-twin is to estimate the transmission rate of each terminal under a given bandwidth and power allocation. The establishment of this sub-twin is important because the system must know whether a terminal can finish communication within the round deadline.

For terminal $n$, let $G_n$ denote the large-scale channel gain, which is determined by path loss~\cite{Wang2025MCOM}. Let $H_n[t]\sim\mathcal{CN}(0,1)$ denote the small-scale fading coefficient in round $t$. If terminal $n$ is allocated bandwidth $b_n[t]$ and transmit power $p_n[t]$, its achievable transmission rate is
\begin{equation}
  R_n[t]
  =
  b_n[t]\log_{2}\!\left(
  1+
  \frac{G_n|H_n[t]|^{2}p_n[t]}{N_0b_n[t]}
  \right),
  \label{eq:sys-R}
\end{equation}
where $N_0$ is the noise power spectral density.

The above rate is then used to estimate how long it takes to upload activations and download gradients during split learning.

\subsection{Training Sub-Twin}
\label{sec:sys-fsl}

\begin{figure}[t]
    \centering
    \includegraphics[width=\linewidth]{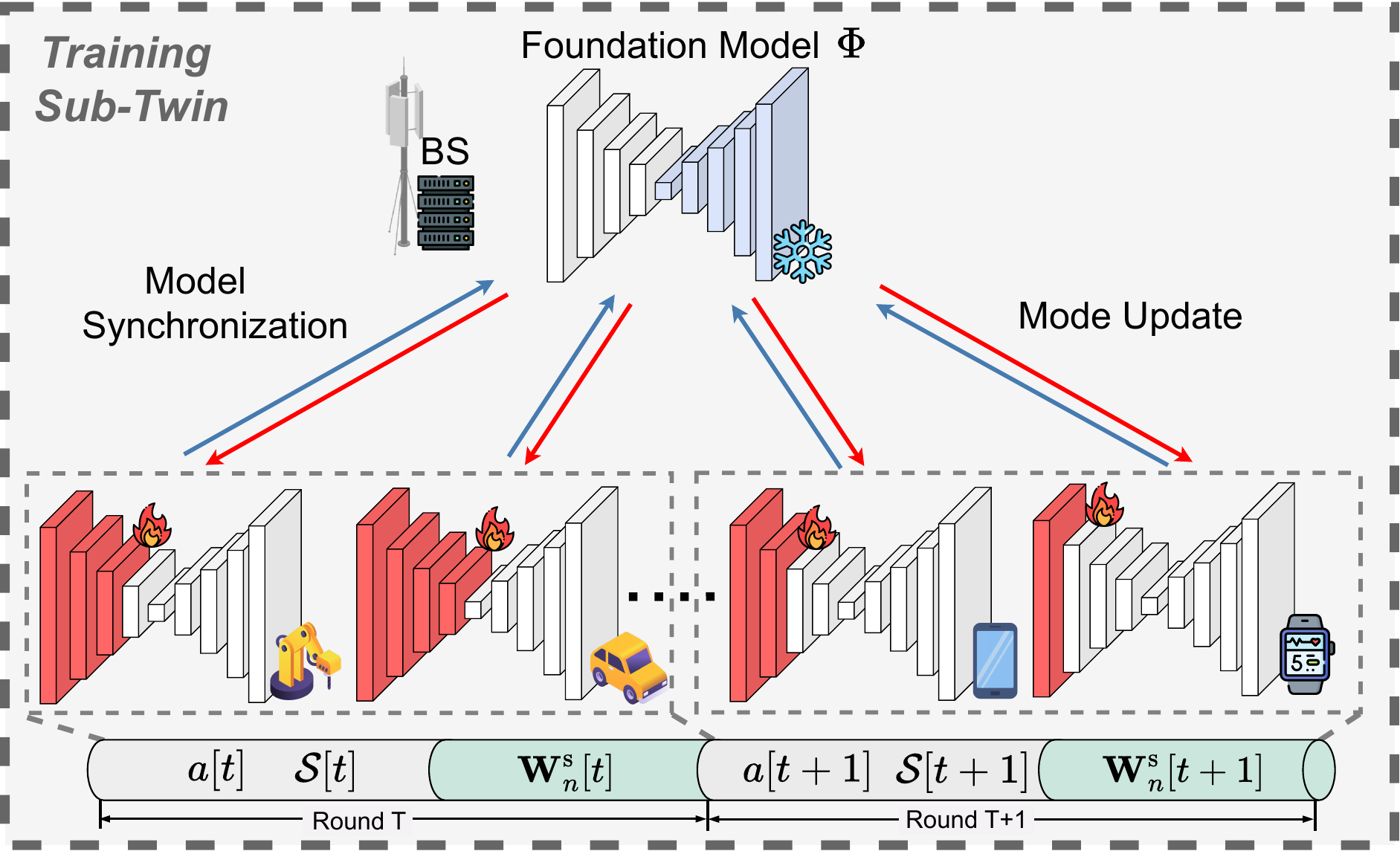}
    \caption{Training sub-twin illustration.}
    \label{fig:training-twin}
\end{figure}

The training sub-twin predicts the effect of split-layer selection, compression, and terminal scheduling on model training. The establishment of this sub-twin is critical because different split layers and compression levels can lead to different training behavior. For example, a deeper split layer may reduce server-side computation but may generate a larger activation payload. A higher compression level may reduce communication delay but may also degrade the model update.

Let $\mathcal{L}$ denote a finite set of admissible split-layer indices, $\mathcal{L}\subset\{1,\ldots,L\}$, where $L$ is the total number of layers in the foundation model $\Phi$.

In round $t$, each scheduled terminal $n\in\mathcal{S}[t]$ chooses a split layer $\ell_n[t]\in\mathcal{L}$. The terminal uses a local mini-batch $\mathcal{B}_n[t]$ with size
$K_n[t]\triangleq|\mathcal{B}_n[t]|$.
Let $\mathbf{X}_n[t]$ denote the input samples in this mini-batch. The terminal processes $\mathbf{X}_n[t]$ through its local part of the model and obtains the split-layer activation
\begin{equation}
  \mathbf{S}_n[t]
  =
  \mathfrak{f}_{\ell_n[t]}
  \bigl(
  \mathbf{X}_n[t];
  \mathbf{W}^{\mathrm{c}}_n[t-1]
  \bigr),
  \label{eq:sys-Sfwd}
\end{equation}
where $\mathbf{W}^{\mathrm{c}}_n[t-1]$ denotes the client-side parameters of terminal $n$ before round $t$.

The terminal compresses $\mathbf{S}_n[t]$ and uploads it to the server. The server then runs the remaining layers, computes the training loss, updates the server-side parameters, and sends the split-layer gradient back to the terminal. The terminal uses this returned gradient to update its local parameters.

Let $\mathbf{W}^{\mathrm{s}}[t]$ denote the server-side parameters after round $t$. Let $\mathbf{W}[t]$ denote all trainable parameters, including both client-side and server-side parameters. For terminal $n$, let $\mathcal{J}_n[t]$ denote the empirical loss associated with its mini-batch in round $t$.

The training loss over all scheduled terminals is defined as
\begin{equation}
    \mathcal{J}[t]
    =
    \sum_{n\in\mathcal{S}[t]}
    \omega_n[t]\mathcal{J}_n[t].
    \label{eq:training-loss}
\end{equation}
Here, $\omega_n[t]=
    \frac{K_n[t]}
    {\sum_{j\in\mathcal{S}[t]}K_j[t]}$ is the mini-batch-size weight of terminal $n$ among all scheduled terminals.

The aggregated server-side gradient is
\begin{equation}
    \mathbf{G}^{\mathrm{s}}[t]
    =
    \sum_{n\in\mathcal{S}[t]}
    \omega_n[t]\mathbf{G}^{\mathrm{s}}_n[t],
    \label{eq:server-gradient}
\end{equation}
where $\mathbf{G}^{\mathrm{s}}_n[t]$ is the gradient of $\mathcal{J}_n[t]$ with respect to the server-side parameters $\mathbf{W}^{\mathrm{s}}[t-1]$.

The server-side model update is
\begin{equation}
  \mathbf{W}^{\mathrm{s}}[t]
  =
  \mathbf{W}^{\mathrm{s}}[t-1]
  -
  \eta^{\mathrm{s}}\mathbf{G}^{\mathrm{s}}[t],
  \label{eq:sys-upd-s}
\end{equation}
where $\eta^{\mathrm{s}}$ is the server-side learning rate.

For each scheduled terminal $n\in\mathcal{S}[t]$, the client-side update is
\begin{equation}
  \mathbf{W}^{\mathrm{c}}_n[t]
  =
  \mathbf{W}^{\mathrm{c}}_n[t-1]
  -
  \eta^{\mathrm{c}}\mathbf{G}^{\mathrm{c}}_n[t],
  \label{eq:sys-upd-c}
\end{equation}
where $\eta^{\mathrm{c}}$ is the client-side learning rate and $\mathbf{G}^{\mathrm{c}}_n[t]$ is the gradient back-propagated from the split layer to the client-side model.

If terminal $n$ is not scheduled in round $t$, its client-side parameters remain unchanged:
\begin{equation}
  \mathbf{W}^{\mathrm{c}}_n[t]
  =
  \mathbf{W}^{\mathrm{c}}_n[t-1],
  \qquad n\notin\mathcal{S}[t].
  \label{eq:sys-upd-unscheduled}
\end{equation}

The training sub-twin uses these update rules to estimate how much model improvement can be obtained by a candidate action $\mathbf{a}[t]$.

\subsection{Task Sub-Twin}
\label{sec:sys-task}

\begin{figure}[t]
    \centering
    \includegraphics[width=\linewidth]{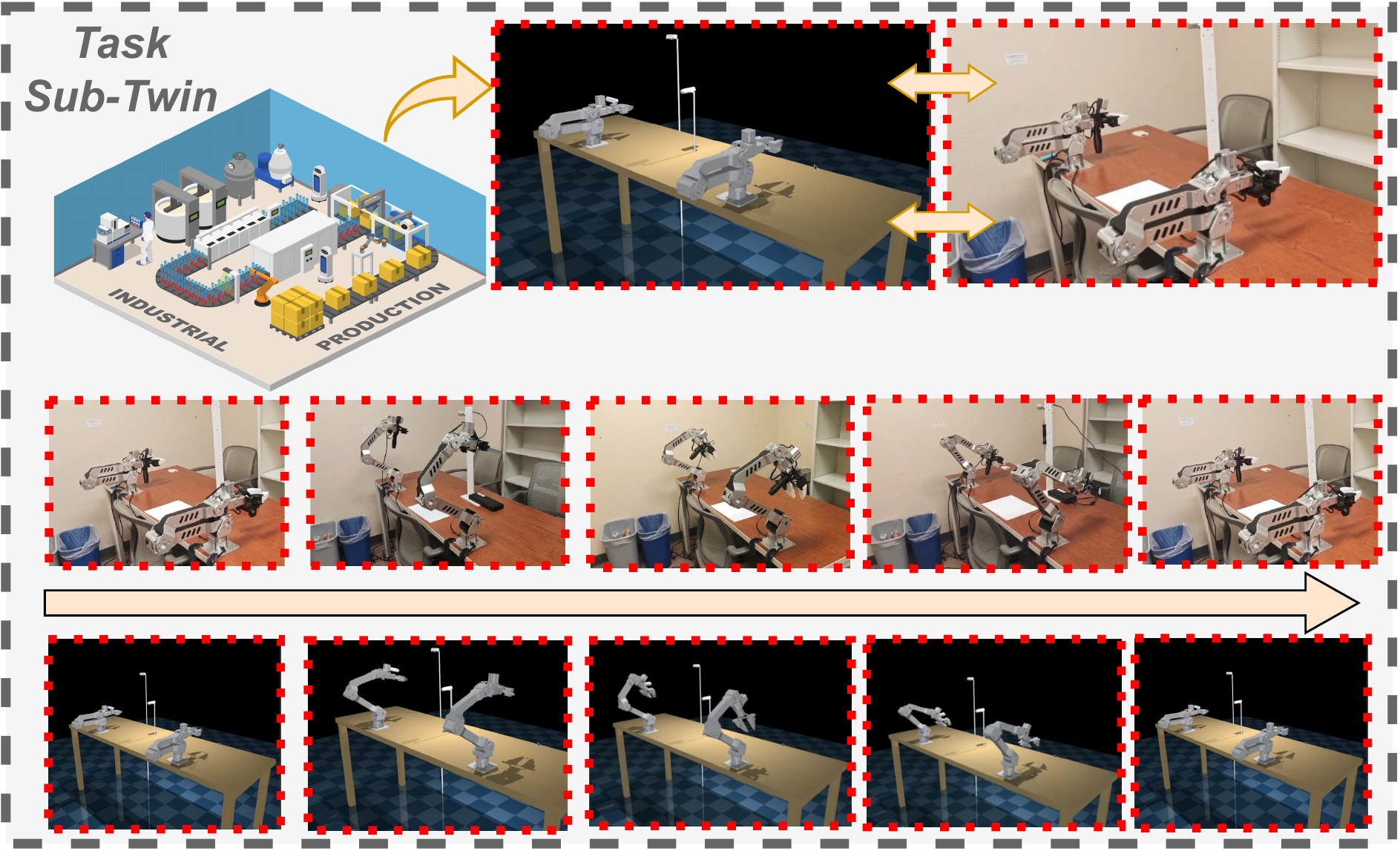}
    \caption{Task sub-twin illustration.}
    \label{fig:task-twin}
\end{figure}

The task sub-twin estimates how well the current model performs in the target task. This is different from the training loss. A model may have a lower training loss, but the main goal is to improve task success in the deployment environment.

Let $\mathcal{E}$ denote a held-out set of evaluation episodes sampled from the deployment environment. For each episode $\tau\in\mathcal{E}$, define $\xi(\mathbf{W},\tau)\in\{0,1\}$
as the task success indicator. Specifically, $\xi(\mathbf{W},\tau)=1$ if the model with parameter $\mathbf{W}$ successfully completes episode $\tau$, and $\xi(\mathbf{W},\tau)=0$ otherwise.

The task success rate at round $t$ is
\begin{equation}
  \Gamma[t]
  =
  \frac{1}{|\mathcal{E}|}
  \sum_{\tau\in\mathcal{E}}
  \xi\bigl(\mathbf{W}[t],\tau\bigr).
  \label{eq:sys-Gamma}
\end{equation}
The task-success improvement in round $t$ is
\begin{equation}
    \Delta\Gamma[t]
    \triangleq
    \Gamma[t]-\Gamma[t-1].
    \label{eq:sys-delta-Gamma}
\end{equation}

In practice, computing $\Gamma[t]$ in every training round is expensive because it requires running many evaluation episodes. Therefore, the task sub-twin evaluates the model in a virtual version of the deployment environment~\cite{Liu2023NeurIPS}. It returns an estimated task success rate $\widehat{\Gamma}[t]$, which provides frequent task-level feedback without requiring physical evaluation in every round.

\subsection{Problem Formulation}
\label{sec:prob}

We now formulate the control problem. The BS must decide which terminals to schedule, how much bandwidth and power to allocate, where to split the model, and how much compression to apply. These decisions affect communication delay, computation delay, energy consumption, memory usage, training progress, and final task performance.

We first define the delay, energy, and memory models used in the constraints.

Let $\psi(\ell)$ denote the per-sample size of the split-layer activation when the model is split at layer $\ell$. Let $\varphi(\ell)$ denote the per-sample computational workload of the client-side sub-model up to layer $\ell$.

For scheduled terminal $n\in\mathcal{S}[t]$, the one-way transmission delay is
\begin{equation}
  \delta^{\mathrm{tx}}_n[t]
  =
  \frac{
  K_n[t]\psi(\ell_n[t])(1-q_n[t])
  }
  {R_n[t]}.
  \label{eq:sys-T}
\end{equation}
This delay applies to both uplink activation transmission and downlink gradient transmission because the split-layer gradient has the same dimension as the split-layer activation.

The on-device computation delay is
\begin{equation}
  \delta^{\mathrm{cp}}_n[t]
  =
  \frac{
  K_n[t]\varphi(\ell_n[t])
  }
  {\nu_n g_n},
  \qquad n\in\mathcal{S}[t],
  \label{eq:sys-Tp}
\end{equation}
where $\nu_n$ is the number of operations per CPU cycle and $g_n$ is the processor frequency of terminal $n$.

A round is completed only after every scheduled terminal finishes its local computation, uploads its activation, and receives the corresponding gradient. Therefore, the round completion time is determined by the slowest scheduled terminal:
\begin{equation}
  \delta[t]
  =
  \max_{n\in\mathcal{S}[t]}
  \left(
  \delta^{\mathrm{cp}}_n[t]
  +
  2\delta^{\mathrm{tx}}_n[t]
  \right).
  \label{eq:sys-TC}
\end{equation}

The energy consumption of scheduled terminal $n$ in round $t$ is
\begin{equation}
  E_n[t]
  =
  \kappa_n g_n^{2}K_n[t]\varphi(\ell_n[t])
  +
  p_n[t]\delta^{\mathrm{tx}}_n[t],
  \label{eq:sys-E}
\end{equation}
where $\kappa_n$ is a hardware-dependent coefficient~\cite{Yang2021TWC}. The first term is the local computation energy, and the second term is the communication energy for uploading the split-layer activation.

Let $m_n(\ell)$ denote the memory required by terminal $n$ when the split layer is $\ell$. This memory includes the client-side model, optimizer state, and cached activations. The memory usage must not exceed the device memory budget $M_n^{\max}$.

At each round $t$, the BS chooses an action $\mathbf{a}[t]$ from a feasible action set $\mathcal{A}[t]$. The feasible set is determined by deadline, bandwidth, power, memory, compression, and scheduling constraints. Let $\delta^{\max}$ denote the per-round deadline and $B^{\max}$ denote the total bandwidth budget. The objective is to maximize the discounted cumulative task-success improvement:
\begin{equation}
\begin{aligned}
  (\mathrm{P_1}):\quad
  \max_{\{\mathbf{a}[t]\}_{t\in\mathcal{T}}}\quad
  &
  \mathbb{E}\!\left[
  \sum_{t\in\mathcal{T}}
  \gamma^{t-1}\Delta\Gamma[t]
  \right]
  \\
  \mathrm{s.t.}\quad
  &(\mathrm{C}_1):\;
  \delta^{\mathrm{cp}}_n[t]+2\delta^{\mathrm{tx}}_n[t]
  \le \delta^{\max},
  && \forall n\in\mathcal{S}[t],\; t\in\mathcal{T},
  \\
  &(\mathrm{C}_2):\;
  m_n(\ell_n[t])\le M_n^{\max},
  && \forall n\in\mathcal{S}[t],\; t\in\mathcal{T},
  \\
  &(\mathrm{C}_3):\;
  \sum_{n\in\mathcal{N}}x_n[t]b_n[t]\le B^{\max},
  && \forall t\in\mathcal{T},
  \\
  &(\mathrm{C}_4):\;
  0\le p_n[t]\le x_n[t]P^{\max},
  && \forall n\in\mathcal{N},\; t\in\mathcal{T},
  \\
  &(\mathrm{C}_5):\;
  0\le b_n[t]\le x_n[t]B^{\max},
  && \forall n\in\mathcal{N},\; t\in\mathcal{T},
  \\
  &(\mathrm{C}_6):\;
  0\le q_n[t]\le q^{\max},
  && \forall n\in\mathcal{S}[t],\; t\in\mathcal{T},
  \\
  &(\mathrm{C}_7):\;
  \ell_n[t]\in\mathcal{L},\quad x_n[t]\in\{0,1\},
  && \forall n\in\mathcal{N},\; t\in\mathcal{T},
  \\
  &(\mathrm{C}_8):\;
  \sum_{n\in\mathcal{N}}x_n[t]\ge 1,
  && \forall t\in\mathcal{T}.
  \label{eq:prob-P}
\end{aligned}
\end{equation}

Constraint $(\mathrm{C}_1)$ ensures that every scheduled terminal completes its computation and communication before the round deadline. Constraint $(\mathrm{C}_2)$ ensures that the selected split layer satisfies the memory limit of the terminal. Constraint $(\mathrm{C}_3)$ limits the total bandwidth used by all scheduled terminals. Constraints $(\mathrm{C}_4)$ and $(\mathrm{C}_5)$ impose power and bandwidth limits. They also force unscheduled terminals to use zero transmit power and zero bandwidth. Constraint $(\mathrm{C}_6)$ limits the compression level. Constraint $(\mathrm{C}_7)$ defines the feasible split-layer and scheduling decisions. Constraint $(\mathrm{C}_8)$ ensures that at least one terminal is scheduled in each round.

Problem $(\mathrm{P_1})$ is difficult for two main reasons. First, the task-success improvement $\Delta\Gamma[t]$ does not have a closed-form expression in terms of the action $\mathbf{a}[t]$. It depends on the entire training trajectory generated by the split-learning updates. Therefore, we use the task sub-twin estimate $\widehat{\Gamma}[t]$ as an online surrogate for evaluating candidate actions.

Second, the decision variables have different roles and time scales. The bandwidth and power variables $(b_n[t],p_n[t])$ mainly respond to fast wireless channel changes. In contrast, the split layer, compression level, and scheduling variables $(\ell_n[t],q_n[t],x_n[t])$ affect the slower evolution of training and task performance. These coupled decisions make the problem challenging.

To address these challenges, we solve the problem round by round. In each round, the cross-domain digital twin $\widehat{\Omega}[t]$ is used as an evaluation proxy to estimate the effect of candidate actions before applying them to the physical system. The proposed method is presented in the next section.
\section{TiLP: Twin-in-the-Loop Planner}
\label{sec:method}

We design TiLP, a twin-in-the-loop planner for solving the problem $(\mathrm{P_1})$. The key idea is that, instead of testing many actions directly on the physical system, TiLP first evaluates candidate actions inside the digital twin $\widehat{\Omega}[t]$. The best action is then applied to the real edge AI system.

At each training round, TiLP treats the control process as a finite-horizon Markov decision process (MDP). The system state summarizes the current wireless condition, learning status, and task-level performance. The action is the system-level control vector $\mathbf{a}[t] =
    \{\mathbf{a}_n[t]\}_{n\in\mathcal{N}},
$
where
$
    \mathbf{a}_n[t]
    =
    \bigl(b_n[t],p_n[t],\ell_n[t],q_n[t],x_n[t]\bigr).
$
TiLP uses the CEM (Cross-Entropy Method)~\cite{Rubinstein1999} to search for a good action sequence inside the digital twin. After simulating several candidate action sequences, TiLP applies only the first action of the best sequence to the physical system. This receding-horizon design allows the planner to adapt to new wireless, learning, and task observations after every round.

\begin{table}[t]
\centering
\setlength{\tabcolsep}{4pt}
\renewcommand{\arraystretch}{1.15}
\caption{Inputs, outputs, and calibration signals of the three sub-twins.}
\label{tab:subtwin-summary}
\resizebox{\columnwidth}{!}{%
\begin{tabular}{@{}l|l|l|l@{}}
\toprule
\textbf{Sub-twin} & \textbf{Input} & \textbf{Output} & \textbf{Calibration signal} \\
\midrule
Network  & $o^{\mathrm{net}}[t],\,b_n[t],\,p_n[t]$
         & $\widehat{R}_n[t],\,\widehat{\delta}[t],\,\widehat{E}_n[t]$
         & $\delta[t],\,E_n[t]$ \\
Training & $o^{\mathrm{tr}}[t],\,\ell_n[t],\,q_n[t],\,x_n[t]$
         & $\Delta\widehat{\mathcal{J}}[t]$
         & $\Delta\mathcal{J}[t]$ \\
Task     & $o^{\mathrm{task}}[t],\,\mathbf{W}[t]$
         & $\widehat{\Gamma}[t]$
         & $\Gamma[t]$ \\
\bottomrule
\end{tabular}}
\end{table}

The digital twin consists of three sub-twins:
\[
    \widehat{\Omega}[t]
    =
    \bigl(
    \widehat{\Omega}^{\mathrm{net}}[t],
    \widehat{\Omega}^{\mathrm{tr}}[t],
    \widehat{\Omega}^{\mathrm{task}}[t]
    \bigr).
\]
Table~\ref{tab:subtwin-summary} summarizes the inputs, outputs, and calibration signals of the three sub-twins. The network sub-twin predicts communication latency and energy. The training sub-twin predicts one-round training progress. The task sub-twin predicts task success. To keep these sub-twins accurate, TiLP calibrates them at three different time scales: every round, every aggregation interval, and every task-evaluation interval.


\subsection{Cross-Domain Sub-Twins}
\label{sec:method-wm}

The digital twin $\widehat{\Omega}[t]$ takes the current state $s[t]$ and a candidate action $\mathbf{a}[t]$ as input. It then predicts the quantities needed to compute the reward in Eq.~\eqref{eq:method-reward}. Throughout this section, hatted variables denote values predicted by the digital twin.

\subsubsection{Network Sub-Twin}

The network sub-twin $\widehat{\Omega}^{\mathrm{net}}[t]$ predicts the communication outcome of a candidate bandwidth and power allocation. For each terminal $n$, it takes the wireless observation and the resource allocation pair $(b_n[t],p_n[t])$ as input. It then predicts the transmission rate $\widehat{R}_n[t]$, the round latency $\widehat{\delta}[t]$, and the terminal energy consumption $\widehat{E}_n[t]$ by using the models in Eqs.~\eqref{eq:sys-R}, \eqref{eq:sys-TC}, and \eqref{eq:sys-E}. 

The network sub-twin has calibration parameters $\boldsymbol{\Psi}^{\mathrm{net}}[t]$. These parameters capture modeling errors in the wireless and hardware models, such as imperfect fading statistics and inaccurate energy coefficients. After each physical round, TiLP compares the predicted latency and energy with the realized latency and energy. The network sub-twin is then updated by reducing the residual between
$
    (\widehat{\delta}[t],\{\widehat{E}_n[t]\}_{n\in\mathcal{N}})
$
and
$
    (\delta[t],\{E_n[t]\}_{n\in\mathcal{N}}).
$
This calibration loop runs once per round.

\subsubsection{Training Sub-Twin}

The training sub-twin $\widehat{\Omega}^{\mathrm{tr}}[t]$ predicts how much training improvement a candidate split-learning decision can produce. Its input includes the scheduled-terminal set $\mathcal{S}[t]$, the split layers $\{\ell_n[t]\}$, the compression levels $\{q_n[t]\}$, and the current learning observation.

The output of the training sub-twin is the predicted one-round loss decrease, denoted by
$
    \Delta\widehat{\mathcal{J}}[t].
$
The prediction is produced by a lightweight parametric model with calibration parameters $\boldsymbol{\Psi}^{\mathrm{tr}}[t]$. These parameters are updated every aggregation interval. Specifically, after every $F$ rounds, TiLP uses a sliding window of recently observed training losses to fit the training sub-twin. The fitting target is the residual between the predicted loss decrease $\Delta\widehat{\mathcal{J}}[\tau]$ and the realized loss decrease $\Delta\mathcal{J}[\tau]$ over recent rounds $\tau$.

\subsubsection{Task Sub-Twin}

The task sub-twin $\widehat{\Omega}^{\mathrm{task}}[t]$ predicts the task success rate of the current model. Its output is denoted by
$
    \widehat{\Gamma}[t].
$
During planning, TiLP uses this predicted task success rate to estimate the task-level benefit of a candidate action. During evaluation, the current model $\mathbf{W}[t]$ is rolled out in the task environment, such as the LIBERO manipulation benchmark~\cite{Liu2023NeurIPS}, to obtain the realized success rate $\Gamma[t]$ according to Eq.~\eqref{eq:sys-Gamma}. The task sub-twin is then updated by reducing the residual between $\widehat{\Gamma}[t]$ and $\Gamma[t]$.

This calibration loop runs every $E_A$ rounds because task evaluation is more expensive than measuring communication latency or training loss.

\subsection{MDP Formulation}
\label{sec:method-mdp}

We formulate the online decision process as a finite-horizon MDP. At each round $t$, the planner observes a state $s[t]$, selects an action $\mathbf{a}[t]$, receives a reward $r[t]$, and moves to the next state $s[t+1]$. The discount factor is $\gamma\in(0,1)$.

The MDP is built around the digital twin. The network and training sub-twins model the one-step evolution of communication and learning states. The task sub-twin does not directly change the model parameters. Instead, it reads the current model state and predicts the task success rate, which contributes to the reward.

\subsubsection{State}

The state contains three observation blocks:
\begin{equation}
  s[t]
  \triangleq
  \bigl(
  o^{\mathrm{net}}[t],
  o^{\mathrm{tr}}[t],
  o^{\mathrm{task}}[t]
  \bigr).
  \label{eq:method-state}
\end{equation}

The network observation $o^{\mathrm{net}}[t]$ includes the large-scale channel gains $\{G_n\}_{n\in\mathcal{N}}$, the instantaneous fading powers $\{|H_n[t]|^2\}_{n\in\mathcal{N}}$, and the previous-round bandwidth allocation $\{b_n[t-1]\}_{n\in\mathcal{N}}$.

The training observation $o^{\mathrm{tr}}[t]$ includes the current training loss $\mathcal{J}[t-1]$ and the previous-round client-side gradient norms
$
    \{\|\mathbf{G}^{\mathrm{c}}_n[t-1]\|\}_{n\in\mathcal{N}}.
$
These quantities summarize the current learning status of the model.

The task observation $o^{\mathrm{task}}[t]$ includes the latest task success estimate $\widehat{\Gamma}[t-1]$. This term informs the planner whether recent training rounds have improved task performance.

\subsubsection{Action}

The action is the system-level control vector
$
    \mathbf{a}[t]
    =
    \{\mathbf{a}_n[t]\}_{n\in\mathcal{N}},
    \label{eq:method-action}
$
where
$
    \mathbf{a}_n[t]
    =
    \bigl(
    b_n[t],
    p_n[t],
    \ell_n[t],
    q_n[t],
    x_n[t]
    \bigr).
$
The action contains both continuous and discrete variables. The bandwidth $b_n[t]$, transmit power $p_n[t]$, and compression level $q_n[t]$ are continuous. The split layer $\ell_n[t]\in\mathcal{L}$ is categorical. The scheduling variable $x_n[t]\in\{0,1\}$ is binary.

The feasible action set $\mathcal{A}[t]$ is determined by the physical constraints in $(\mathrm{P_1})$, including deadline, memory, bandwidth, power, compression, split-layer, and scheduling constraints.

\subsubsection{Reward}

The reward should encourage three desirable behaviors: improving task success, reducing communication cost, and avoiding constraint violations. Therefore, we define the per-round reward as
\begin{equation}
  r[t]
  =
  r_{\mathrm{task}}[t]
  +
  r_{\mathrm{comm}}[t]
  -
  r_{\mathrm{pen}}[t].
  \label{eq:method-reward}
\end{equation}

The first term measures task-level improvement:
\begin{equation}
    r_{\mathrm{task}}[t]
    =
    \Delta\widehat{\Gamma}[t]
    =
    \widehat{\Gamma}[t]-\widehat{\Gamma}[t-1].
    \label{eq:method-r-task}
\end{equation}
This term corresponds to the objective of $(\mathrm{P_1})$, but it uses the task sub-twin estimate during online planning.

The second term rewards communication efficiency:
\begin{equation}
  r_{\mathrm{comm}}[t]
  =
  w_{\delta}
  \left(
  1-\frac{\widehat{\delta}[t]}{\delta^{\max}}
  \right)
  +
  w_E
  \left(
  1-\frac{\sum_{n\in\mathcal{S}[t]}\widehat{E}_n[t]}{E^{\max}}
  \right),
  \label{eq:method-r-comm}
\end{equation}
where $w_{\delta}>0$ and $w_E>0$ are weighting coefficients. The constant
\begin{equation}
    E^{\max}
    \triangleq
    NP^{\max}\delta^{\max}
\end{equation}
is used to normalize the round energy. This communication reward is larger when the predicted latency is well below the deadline and the predicted energy consumption is small.

The third term penalizes violations of major physical constraints:
\begin{equation}
    r_{\mathrm{pen}}[t]
    =
    w_{\mathrm{pen}}\mathcal{V}[t],
    \label{eq:method-r-pen}
\end{equation}
where $w_{\mathrm{pen}}>0$ is the penalty weight and $\mathcal{V}[t]$ is the aggregate constraint-violation function:
\begin{equation}
\begin{split}
  \mathcal{V}[t]
  &=
  \sum_{n\in\mathcal{S}[t]}
  \max\left(
  0,
  \widehat{\delta}^{\mathrm{cp}}_n[t]
  +
  2\widehat{\delta}^{\mathrm{tx}}_n[t]
  -
  \delta^{\max}
  \right)
  \\
  &\quad+
  \sum_{n\in\mathcal{S}[t]}
  \max\left(
  0,
  m_n(\ell_n[t])-M_n^{\max}
  \right)
  \\
  &\quad+
  \max\left(
  0,
  \sum_{n\in\mathcal{N}}x_n[t]b_n[t]-B^{\max}
  \right).
\end{split}
\label{eq:method-violation}
\end{equation}
The first term penalizes deadline violation. The second term penalizes memory violation. The third term penalizes bandwidth-budget violation. If all these constraints are satisfied, then $\mathcal{V}[t]=0$.

The remaining constraints are enforced directly during action generation. Power and bandwidth samples are clipped to feasible intervals. Split-layer samples are projected onto $\mathcal{L}$. All-zero schedules are rejected so that at least one terminal is selected in each round.

The MDP objective is
\begin{equation}
    \max_{\pi}
    \mathbb{E}_{\pi}
    \left[
    \sum_{t\in\mathcal{T}}
    \gamma^{t-1}r[t]
    \right].
    \label{eq:method-mdp-objective}
\end{equation}
Compared with $(\mathrm{P_1})$, this MDP objective keeps the task-success improvement as the main goal while adding communication efficiency and soft constraint handling.

\subsection{Decision-Making Process}
\label{sec:method-plan}

At each round $t$, TiLP selects the action $\mathbf{a}^{\star}[t]$ by planning inside the digital twin. The planner looks ahead for $K$ future rounds. It evaluates many candidate $K$-step action sequences and selects the one with the highest predicted return. Only the first action of the best sequence is applied to the physical system. At the next round, the planner observes the new state and repeats the process.

TiLP uses CEM to search over the mixed continuous-discrete action space. CEM maintains a sampling distribution over candidate action sequences. At the beginning of round $t$, this distribution is initialized from a learned actor policy $\pi_{\theta}(\mathbf{a}\mid s[t])$. The actor provides a good initial guess, while CEM improves the action by explicit planning inside the twin.

For each CEM iteration, TiLP samples a population of candidate action sequences:
$
    \widehat{\mathbf{a}}[t],
    \widehat{\mathbf{a}}[t+1],
    \ldots,
    \widehat{\mathbf{a}}[t+K-1].
$
Each sequence is rolled out inside $\widehat{\Omega}[t]$. The rollout produces predicted rewards
$
    \widehat{r}[t],
    \widehat{r}[t+1],
    \ldots,
    \widehat{r}[t+K-1]
$
and a predicted terminal state $\widehat{s}[t+K]$.

The score of a candidate sequence is
\begin{equation}
  \widehat{J}
  =
  \sum_{k=0}^{K-1}
  \gamma^k\widehat{r}[t+k]
  +
  \gamma^K V(\widehat{s}[t+K]),
  \label{eq:method-J}
\end{equation}
where the terminal value is approximated by
\begin{equation}
  V(s)
  \triangleq
  \mathbb{E}_{\mathbf{a}\sim\pi_{\theta}(\cdot\mid s)}
  \left[
  Q_{\phi}(s,\mathbf{a})
  \right].
  \label{eq:method-V}
\end{equation}
Here, $Q_{\phi}(s,\mathbf{a})$ is a learned critic that estimates the long-term value of taking action $\mathbf{a}$ at state $s$.

After scoring all candidate sequences, CEM keeps the top-performing sequences, called elites, and refits its sampling distribution to these elites. This process repeats for a fixed number of iterations. After the final iteration, TiLP selects the first action of the best sequence:
$
    \mathbf{a}^{\star}[t]
    =
    \widehat{\mathbf{a}}^{\star}[t].
$
The look-ahead horizon $K$ is kept small because long rollouts can accumulate twin prediction errors.

The actor policy factorizes across terminals:
\begin{equation}
  \pi_{\theta}(\mathbf{a}\mid s)
  =
  \prod_{n\in\mathcal{N}}
  \pi_{\theta}(\mathbf{a}_n\mid s).
  \label{eq:method-pi}
\end{equation}
For terminal $n$, the continuous variables $(b_n,p_n,q_n)$ are sampled from a tanh-squashed Gaussian distribution~\cite{Jennifer2018PMLR}. The split-layer variable $\ell_n$ is sampled from a Gumbel-softmax categorical distribution over $\mathcal{L}$~\cite{Eric2017ICLR}. The scheduling variable $x_n$ is sampled from a Bernoulli distribution.

The critic $Q_{\phi}(s,\mathbf{a})$ is implemented as a feed-forward neural network. It takes the concatenated state-action vector as input and outputs a scalar value estimate. The actor and critic are trained with soft actor-critic (SAC)~\cite{Haarnoja2018ICML}. Training uses both real transitions collected from the physical system and imagined transitions generated by rolling out the actor inside the digital twin~\cite{Janner2019NeurIPS}.

The above process creates a closed learning loop. The actor and critic guide CEM by providing the initial sampling distribution and terminal value estimate. CEM selects stronger actions, and these executed actions generate better replay data for training the actor and critic. As the actor and critic improve, they provide better guidance for future CEM planning.

\subsection{Algorithm and Multi-Scale Calibration}
\label{sec:method-proc}

Algorithm~\ref{alg:tilp} summarizes TiLP over $T$ rounds. Each round has three phases.

In Phase I, TiLP observes the current state and runs CEM inside the digital twin to select the action $\mathbf{a}^{\star}[t]$.

In Phase II, the selected action is executed on the physical FSL system. The scheduled terminals perform client-side forward and backward propagation, while the server performs server-side forward and backward propagation. The system then observes the realized latency, energy, training loss, and reward.

In Phase III, TiLP updates the digital twin and the actor-critic models. The three sub-twins are calibrated at different frequencies because they model processes that evolve at different time scales.

\begin{itemize}
    \item \textbf{Loop 1: Network calibration.} This loop runs every round. It updates $\boldsymbol{\Psi}^{\mathrm{net}}[t]$ by reducing the residual between predicted and realized latency and energy.

    \item \textbf{Loop 2: Training calibration.} This loop runs every $F$ rounds. It updates $\boldsymbol{\Psi}^{\mathrm{tr}}[t]$ by reducing the residual between predicted and realized loss decrease over a recent observation window.

    \item \textbf{Loop 3: Task calibration.} This loop runs every $E_A$ rounds. It evaluates the current model in the task environment, obtains the realized success rate $\Gamma[t]$, and updates $\boldsymbol{\Psi}^{\mathrm{task}}[t]$ by reducing the residual between $\widehat{\Gamma}[t]$ and $\Gamma[t]$.
\end{itemize}

\begin{algorithm}[t]
\small
\caption{TiLP: Twin-in-the-Loop Planner}
\label{alg:tilp}
\textbf{Input:} Initial model $\mathbf{W}[0]$; digital twin $\widehat{\Omega}[0]$ with parameters $\boldsymbol{\Psi}^{\mathrm{net}}[0]$, $\boldsymbol{\Psi}^{\mathrm{tr}}[0]$, and $\boldsymbol{\Psi}^{\mathrm{task}}[0]$; actor $\pi_{\theta}$; critic $Q_{\phi}$ with target critic $Q_{\phi^-}$; planning horizon $K$; aggregation interval $F$; evaluation interval $E_A$; CEM population size $Y_1$; number of elites $Y_2$; CEM iterations $I$; imagination rollout length $U$.
\begin{algorithmic}
  \STATE Initialize $\pi_{\theta}$ and $Q_{\phi}$; initialize real and imagined replay buffers

  \FOR{$t=1,\ldots,T$}

    \colorbox{white}{\parbox{\dimexpr\columnwidth-2\fboxsep-20pt\relax}{\STATE \gray{$\triangleright$ \textit{Phase I: Plan inside the digital twin}}}}
    \colorbox{rgb:red!2,65;green!30,60;blue!20,125}{\parbox{\dimexpr\columnwidth-2\fboxsep-20pt\relax}{\vbox{
    \STATE Assemble the state $s[t]$ according to Eq.~\eqref{eq:method-state}
    \STATE Initialize the CEM sampling distribution $\mu$ from $\pi_{\theta}(\cdot\mid s[t])$

    \FOR{$i=1,\ldots,I$}
        \STATE Sample $Y_1$ candidate $K$-step action sequences from $\mu$
        \STATE Roll out each candidate sequence inside $\widehat{\Omega}[t]$
        \STATE Score each sequence by Eq.~\eqref{eq:method-J}
        \STATE Refit $\mu$ to the top $Y_2$ elite sequences
    \ENDFOR

    \STATE Set $\mathbf{a}^{\star}[t]$ as the first action of the highest-scoring sequence
    }}}

    \colorbox{white}{\parbox{\dimexpr\columnwidth-2\fboxsep-20pt\relax}{\STATE \gray{$\triangleright$ \textit{Phase II: Execute one physical FSL round}}}}
    \colorbox{rgb:red!2,65;green!30,90;blue!20,125}{\parbox{\dimexpr\columnwidth-2\fboxsep-20pt\relax}{\vbox{
    \STATE Broadcast $\mathbf{a}^{\star}[t]$ to the terminals
    \STATE Each scheduled terminal $n\in\mathcal{S}[t]$ performs client-side forward propagation according to Eq.~\eqref{eq:sys-Sfwd}
    \STATE The server performs server-side forward propagation, loss computation, and server-side update according to Eq.~\eqref{eq:sys-upd-s}
    \STATE Each scheduled terminal performs client-side back-propagation and updates its local parameters according to Eq.~\eqref{eq:sys-upd-c}
    \STATE Observe $\delta[t]$, $\{E_n[t]\}_{n\in\mathcal{S}[t]}$, and the realized training loss
    \STATE Compute $r[t]$ according to Eqs.~\eqref{eq:method-reward} to~\eqref{eq:method-violation}
    }}}

    \colorbox{white}{\parbox{\dimexpr\columnwidth-2\fboxsep-20pt\relax}{\STATE \gray{$\triangleright$ \textit{Phase III: Calibrate twins and update the actor-critic models}}}}
    \colorbox{rgb:red!60,100;green!20,90;blue!30,125}{\parbox{\dimexpr\columnwidth-2\fboxsep-20pt\relax}{\vbox{
    \STATE Update $\boldsymbol{\Psi}^{\mathrm{net}}[t]$ using the residual between predicted and realized latency and energy
    \STATE Append $(s[t],\mathbf{a}^{\star}[t],r[t],s[t+1])$ to the real replay buffer
    \STATE Generate a $U$-step imagined rollout under $\pi_{\theta}$ and $\widehat{\Omega}[t]$; append it to the imagined replay buffer
    \STATE Update $\pi_{\theta}$ and $Q_{\phi}$ by SAC using the union of real and imagined replay buffers
    \STATE Soft-update the target critic $Q_{\phi^-}$

    \IF{$t \bmod F = 0$}
        \STATE Aggregate client-side models when aggregation is scheduled
        \STATE Update $\boldsymbol{\Psi}^{\mathrm{tr}}[t]$ using recent loss-reduction residuals
    \ENDIF

    \IF{$t \bmod E_A = 0$}
        \STATE Evaluate $\mathbf{W}[t]$ in the task environment and obtain $\Gamma[t]$ according to Eq.~\eqref{eq:sys-Gamma}
        \STATE Update $\boldsymbol{\Psi}^{\mathrm{task}}[t]$ using the residual between $\widehat{\Gamma}[t]$ and $\Gamma[t]$
    \ENDIF
    }}}

  \ENDFOR

  \STATE \textbf{Output:} Final model $\mathbf{W}[T]$ and task success rate $\Gamma[T]$
\end{algorithmic}
\end{algorithm}
\section{Experimental Results}
\label{sec:exp}

\subsection{Experimental Setup}
\label{sec:exp-setup}

We evaluate TiLP on task-oriented federated split learning over a wireless edge network. The experiments are designed to answer three questions. First, does TiLP improve the final task success rate? Second, does it reach a target success rate faster than existing FSL baselines? Third, does it achieve these gains without excessive latency, energy, or communication overhead?

The experimental setup has two parts. The first part contains fixed system parameters, such as the number of training rounds, deadline, memory budget, and CEM settings. These parameters define one training run. The second part contains time-varying quantities, such as fading, loss evolution, and task success estimates. These quantities make the problem a sequential decision-making problem rather than a static optimization problem.

\subsubsection{Digital-Twin Hardware Testbed}
\label{sec:exp-setup-testbed}

The hardware testbed implements the cross-domain digital twin using three compute modules and a physical wireless-robotic front end. The platform includes one high-performance computer (HPC) with two NVIDIA H100 GPUs, two NVIDIA Jetson AGX modules, four USRP software-defined radios, two HighTorque robotic arms, and two Intel RealSense depth cameras.

The HPC hosts the training sub-twin $\widehat{\Omega}^{\mathrm{tr}}[t]$ and performs the BS-side computation for FSL. Specifically, it runs the server-side forward and backward passes, aggregates gradients, updates model parameters, and generates the calibrated loss-reduction estimates used by Loop 2. AGX2 hosts the network sub-twin $\widehat{\Omega}^{\mathrm{net}}[t]$. It controls three USRP-2900 radios that represent three physical terminals. A USRP-2955, together with Sionna RT running on AGX2, emulates the wireless channel used to transmit split-layer activations and gradients between the terminals and the HPC. The remaining terminals are emulated in Sionna RT on the same module, so that the FSL system can operate at the full value of $N$ specified in Table~\ref{tab:system-params}. This hybrid design keeps Loop 1 grounded in measured channel statistics from the on-air terminals while preserving the scale needed for system-level convergence experiments.

AGX1 hosts the task sub-twin $\widehat{\Omega}^{\mathrm{task}}[t]$. It is connected to two HighTorque robotic arms, one RealSense D435 side-view depth camera, and one RealSense D405 wrist-mounted depth camera. During planning, AGX1 runs the MuJoCo-based LIBERO manipulation simulator to evaluate candidate policy parameters. Periodically, the physical robotic arms and depth cameras provide ground-truth rollout measurements, which are used to calibrate Loop 3.

\begin{figure}[t]
    \centering
    \includegraphics[width=1\linewidth]{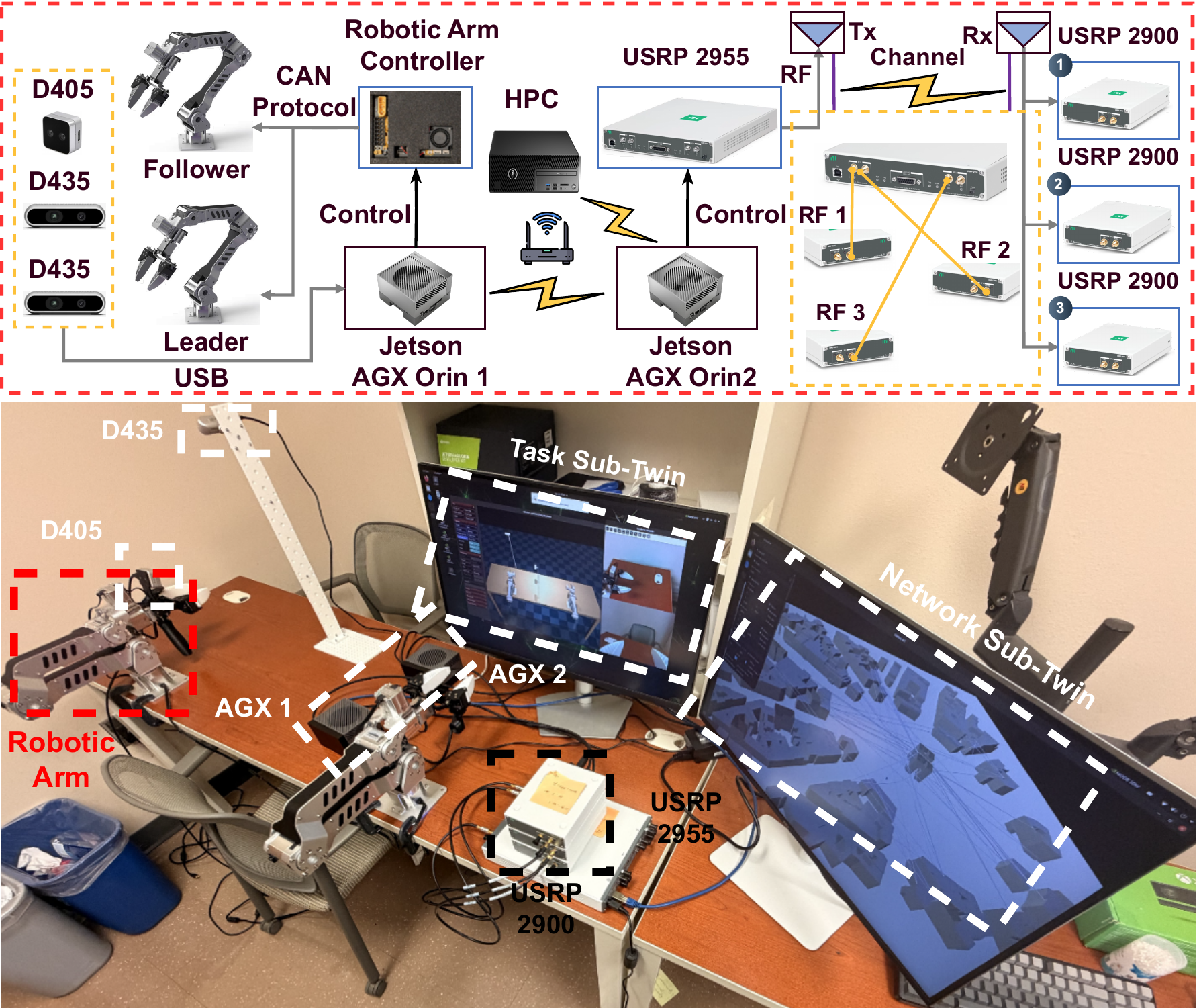}
    \caption{Hardware testbed: HPC (Two H100 GPUs, training sub-twin), AGX2 (network sub-twin, USRP-2900/2955 and Sionna RT), and AGX1 (task sub-twin, HighTorque arms, RealSense D435\,/\,D405, MuJoCo), bridged by gigabit Ethernet.}
    \label{fig:hardware-topology}
\end{figure}

The three compute modules are connected through wired gigabit Ethernet. In each round, split-layer activations and gradients are exchanged between AGX2 and the HPC under the round deadline $\delta^{\max}$. Every $E_A$ rounds, the HPC transfers the current model parameters $\mathbf{W}[t]$ to AGX1 for held-out task-success evaluation according to Eq.~\eqref{eq:sys-Gamma}. Fig.~\ref{fig:hardware-topology} summarizes the hardware topology and data flow.

\subsubsection{System Parameters}
\label{sec:exp-setup-sys}

Table~\ref{tab:system-params} lists the default parameters used in the experiments. Unless otherwise stated, all experiments use these values. In each parameter sweep, only the parameter being studied is changed, while all other parameters remain fixed.

\begin{table}[!t]
\centering
\renewcommand{\arraystretch}{1.15}
\setlength{\tabcolsep}{4pt}
\caption{Default parameter settings.}
\label{tab:system-params}
\small
\begin{tabular}{@{}p{0.15\linewidth}p{0.58\linewidth}p{0.20\linewidth}@{}}
\toprule
\textbf{Symbol} & \textbf{Description} & \textbf{Value} \\
\midrule
$N$                         & Number of terminals                                      & $50$ \\
$T$                         & Total number of training rounds                          & $1000$ \\
$\delta^{\max}$             & Per-round deadline                                       & $5$\,s \\
$B^{\max}$                  & Total bandwidth budget                                   & $100$\,MHz \\
$P^{\max}$                  & Maximum transmit power of each terminal                  & $0.2$\,W \\
$q^{\max}$                  & Maximum payload compression level                        & $0.9$ \\
$M_n^{\max}$                & Memory budget of terminal $n$                            & $8$\,GB \\
$N_0$                       & Noise power spectral density                             & $-174$\,dBm/Hz \\
$\mathcal{L}$               & Set of admissible split-layer indices                    & $\{k_1,\cdots,k_5\}$ \\
$g_n$                       & Processor frequency of terminal $n$                      & $1.5$\,GHz \\
$F$                         & Federated aggregation interval                           & $10$ rounds \\
$E_A$                       & Task-evaluation interval                                 & $50$ rounds \\
$K$                         & CEM planning horizon                                     & $16$ \\
$Y_1,Y_2$                   & CEM population size and number of elites                 & $200,\,25$ \\
$I$                         & CEM iterations per round                                 & $5$ \\
$U$                         & Imagination rollout length                               & $10$ \\
$\gamma$                    & Discount factor                                          & $0.95$ \\
\bottomrule
\end{tabular}
\end{table}

\begin{figure*}[t]
    \centering
    \subfloat[Task success rate]{\includegraphics[width=0.32\linewidth]{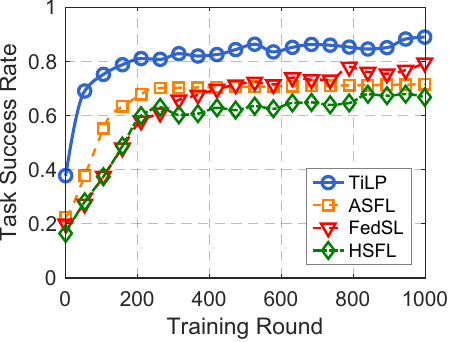}\label{fig:main-success}}
    \hfill
    \subfloat[Round latency]{\includegraphics[width=0.32\linewidth]{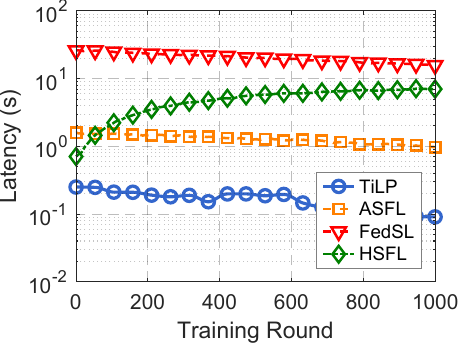}\label{fig:main-latency}}
    \hfill
    \subfloat[Cumulative uplink volume]{\includegraphics[width=0.32\linewidth]{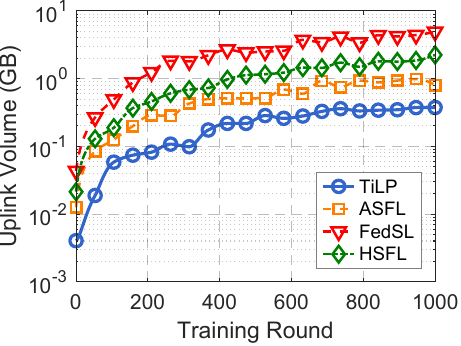}\label{fig:main-volume}}
    \caption{Main convergence comparison between TiLP and representative FSL baselines, including ASFL~\cite{Xu2024TWC}, FedSL~\cite{Ao2025TCE}, and HSFL~\cite{Lin2025TMC_a}.}
    \label{fig:main-convergence}
\end{figure*}

\begin{table*}[t]
\centering
\setlength{\tabcolsep}{6pt}
\renewcommand\arraystretch{1.15}
\caption{Main results of \method{} against eight FSL baselines. \best{Best} in red bold, \second{runner-up} in blue; ``--'' denotes an unreached threshold or an undefined metric.}
\label{tab:main-results}
\resizebox{0.98\linewidth}{!}{
\begin{tabular}{l|c|c|c|c|c|c}
\Xhline{1.2pt}
\textbf{Method} & \textbf{Success (\%)} & \textbf{RTA (60/70/80)} & \textbf{Latency (s)} & \textbf{Energy (norm.)} & \textbf{Volume (GB)} & \textbf{Violation} \\
\Xhline{1.2pt}
FedAvg \venue{AISTATS'17}~\cite{Singh2017PMLR}    & 55.5\std{5.4} & 985 / -- / --       & 23.40\std{0.46} & 77.5\std{2.1}          & 9.12\std{0.24} & 29.50\std{0.05} \\
SplitNN \venue{NeurIPS'18}~\cite{Praneeth2018arXiv} & 58.1\std{4.3} & 820 / -- / --       & 20.15\std{0.31} & 49.7\std{1.2}          & 7.85\std{0.19} & 25.30\std{0.04} \\
SplitFed \venue{AAAI'22}~\cite{Thapa2022AAAI}     & 61.8\std{3.9} & 680 / -- / --        & 18.90\std{0.28} & 41.1\std{0.9}          & 6.95\std{0.17} & 21.80\std{0.03} \\
AdaptSFL \venue{TMC'26}~\cite{Lin2025TMC_b}        & 64.6\std{3.6} & 560 / -- / --       & 18.25\std{0.10} & 23.3\std{0.7}          & 5.80\std{0.16} & 17.10\std{0.03} \\
GA-SFL \venue{TWC'25}~\cite{Liang2026TWC}             & 67.4\std{3.5} & 420 / -- / --       & 16.05\std{0.09} & 19.3\std{0.6}          & 5.22\std{0.14} & 13.75\std{0.02} \\
HSFL \venue{TMC'25}~\cite{Lin2025TMC_a}                 & 68.0\std{2.9} & 210 / -- / --       & 7.34\std{0.04}  & 10.1\std{0.3}          & 2.08\std{0.06}           & 9.99\std{0.01} \\
ASFL \venue{TWC'24}~\cite{Xu2024TWC}                  & 71.5\std{3.2} & 120 / 235 / --  & \second{1.06\std{0.02}} & \second{1.28\std{0.03}} & \second{1.00\std{0.03}} & \second{4.06\std{0.01}} \\
FedSL \venue{TCE'25}~\cite{Ao2025TCE}                 & \second{78.0\std{2.5}} & 220 / 410 / -- & 15.10\std{0.08} & 14.9\std{0.4}          & 4.95\std{0.12}           & 12.83\std{0.01} \\
\hline
\textbf{\method{} (Ours)}                                                & \best{87.5\std{2.0}} & \best{29 / 68 / 144} & \best{0.10\std{0.01}} & \best{1.00} & \best{0.42\std{0.01}} & \best{3.52\std{0.02}} \\
\Xhline{1.2pt}
\end{tabular}}
\end{table*}

\subsubsection{Dynamic Environment}
\label{sec:exp-setup-dyn}

Although the parameters in Table~\ref{tab:system-params} are fixed during one run, the environment changes over time. The fading power $|H_n[t]|^2$ evolves across rounds according to a first-order auto-regressive process. The Doppler shift $f_D$ and temporal correlation coefficient $\rho_H$ control the speed and smoothness of this channel variation.

The learning state also changes over time. The training loss $\mathcal{J}[t]$ and the client-side gradient norms $\{\|\mathbf{G}^{\mathrm{c}}_n[t]\|\}_{n\in\mathcal{N}}$ evolve as the model is trained. Around each aggregation interval, the value of selecting a terminal may change because its local model and gradient contribution may become more or less useful. The task success estimate $\widehat{\Gamma}[t]$ evolves more slowly because task evaluation is performed only every $E_A$ rounds.

These time-varying components make the feasible action set and the reward change from round to round. The scheduling decision at round $t$ also affects the learning state observed at round $t+1$. Therefore, the problem cannot be reduced to a one-shot resource allocation problem. This motivates the receding-horizon planning approach described in Section~\ref{sec:method}.

\subsection{Performance Metrics}
\label{sec:exp-metrics}

We evaluate TiLP using six metrics. The first two measure task capability and convergence speed. The next three measure communication and resource cost. The last one measures constraint violation.

\subsubsection{Final Task Success Rate}

The main performance metric is the final task success rate:
$
    \Gamma[T],
$
which is computed on held-out task episodes using Eq.~\eqref{eq:sys-Gamma}. This metric directly measures whether the final model succeeds in the deployment task.

\subsubsection{Rounds-to-Accuracy}

The rounds-to-accuracy metric measures how quickly a method reaches a target task success rate. For a threshold $\vartheta$, it is defined as
\begin{equation}
  \mathrm{RTA}(\vartheta)
  \triangleq
  \min\{t\in\mathcal{T}: \Gamma[t]\ge\vartheta\}.
  \label{eq:metric-rta}
\end{equation}
We report $\mathrm{RTA}(\vartheta)$ at $\vartheta\in\{0.6,0.7,0.8\}$. A smaller value means that the method reaches the target success rate earlier.

\subsubsection{Cumulative Energy Consumption}

The cumulative terminal energy over the full training horizon is
\begin{equation}
  E_{\Sigma}
  \triangleq
  \sum_{t\in\mathcal{T}}
  \sum_{n\in\mathcal{S}[t]}
  E_n[t],
  \label{eq:metric-energy}
\end{equation}
where $E_n[t]$ is defined in Eq.~\eqref{eq:sys-E}. We report normalized energy, with the TiLP run under the default setting used as the reference value. Thus, a value larger than one means that the method consumes more energy than TiLP.

\subsubsection{Average Round Latency}

The average round latency is
\begin{equation}
  \bar{\delta}
  \triangleq
  \frac{1}{T}
  \sum_{t\in\mathcal{T}}
  \delta[t],
  \label{eq:metric-latency}
\end{equation}
where $\delta[t]$ is the straggler-dominated round latency in Eq.~\eqref{eq:sys-TC}. A smaller $\bar{\delta}$ indicates that the scheduler better controls the slowest scheduled terminal.

\subsubsection{Cumulative Uplink Communication Volume}

The cumulative uplink payload is
\begin{equation}
  V_{\Sigma}
  \triangleq
  \sum_{t\in\mathcal{T}}
  \sum_{n\in\mathcal{S}[t]}
  K_n[t]\psi(\ell_n[t])\bigl(1-q_n[t]\bigr).
  \label{eq:metric-volume}
\end{equation}
This metric captures the combined effect of scheduling, split-layer selection, and compression. Scheduling determines which terminals transmit. The split layer $\ell_n[t]$ determines the activation size $\psi(\ell_n[t])$. The compression level $q_n[t]$ determines the fraction of the payload that is discarded.

\subsubsection{Average Constraint Violation}

We use the average constraint-violation value to measure how closely a method follows the latency, memory, and bandwidth limits:
\begin{equation}
  \bar{\mathcal{V}}
  \triangleq
  \frac{1}{T}
  \sum_{t\in\mathcal{T}}
  \mathcal{V}[t],
  \label{eq:metric-violation}
\end{equation}
where $\mathcal{V}[t]$ is defined in Eq.~\eqref{eq:method-violation}. A smaller $\bar{\mathcal{V}}$ means fewer or smaller constraint violations. If all softened constraints are satisfied in every round, then $\bar{\mathcal{V}}=0$.

\subsection{Main Results}
\label{sec:exp-main}

Fig.~\ref{fig:main-convergence} compares the training trajectories of TiLP and representative FSL baselines. TiLP improves task success faster while keeping latency and uplink volume low. This result shows that the gain is not obtained by simply spending more wireless resources.

Table~\ref{tab:main-results} gives the full comparison across all baselines. TiLP achieves a final task success rate of $87.5\%$, which is $9.5$ percentage points higher than the strongest baseline, FedSL. TiLP also reaches the $60\%$ success threshold in $29$ rounds, compared with $120$ rounds for ASFL and $220$ rounds for FedSL. For communication cost, TiLP keeps the average latency at $0.10$\,s and the cumulative uplink volume at $0.42$\,GB. These results indicate that the twin-based planner selects actions that improve task success while avoiding unnecessary communication.

\begin{figure*}[t]
    \centering
    \subfloat[Task success vs. retention]{\includegraphics[width=0.32\linewidth]{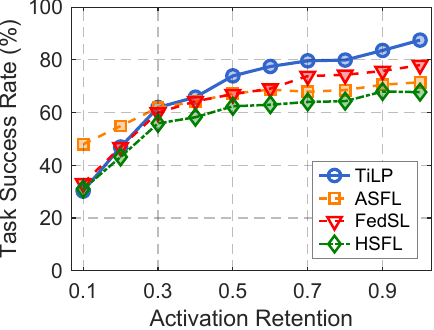}\label{fig:rho-success}}
    \hfill
    \subfloat[Round latency vs. retention]{\includegraphics[width=0.32\linewidth]{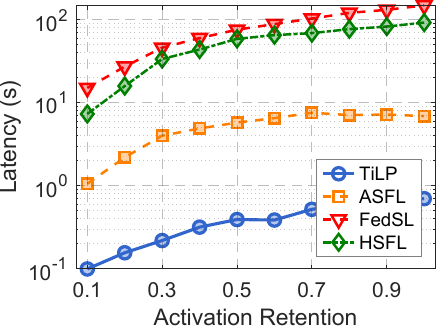}\label{fig:rho-latency}}
    \hfill
    \subfloat[Cumulative uplink volume vs. retention]{\includegraphics[width=0.32\linewidth]{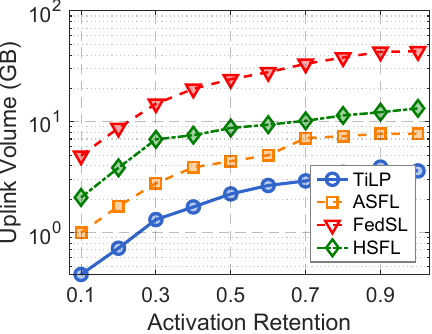}\label{fig:rho-volume}}
    \\
    \subfloat[Task success vs. split-layer strategy]{\includegraphics[width=0.32\linewidth]{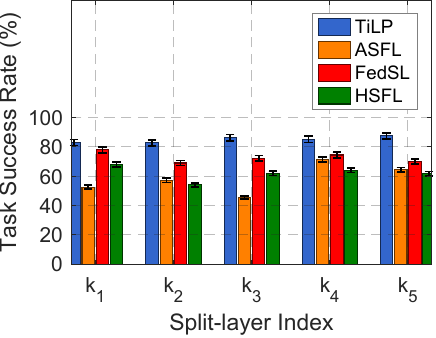}\label{fig:split-success}}
    \hfill
    \subfloat[Round latency vs. split-layer strategy]{\includegraphics[width=0.32\linewidth]{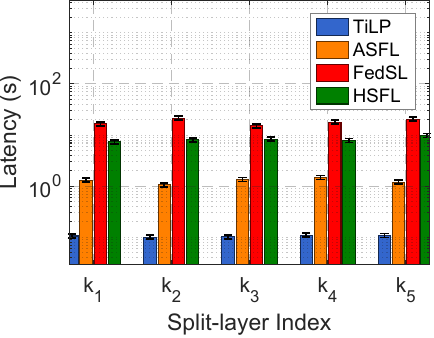}\label{fig:split-latency}}
    \hfill
    \subfloat[Cumulative uplink volume vs. split-layer strategy]{\includegraphics[width=0.32\linewidth]{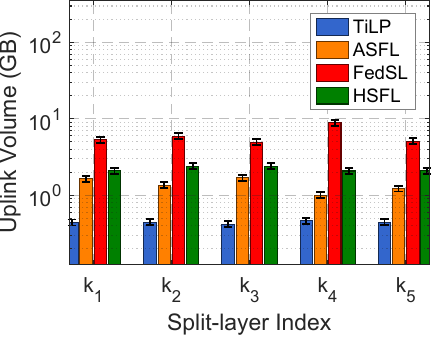}\label{fig:split-volume}}
    \caption{Cross-layer robustness of TiLP. The top row varies the activation retention ratio $1-q_n[t]$. The bottom row varies the split-layer strategy. TiLP is compared with ASFL, FedSL, and HSFL.}
    \label{fig:crosslayer-ablation}
\end{figure*}

\begin{figure*}[t]
    \centering
    \subfloat[Task success rate]{\includegraphics[width=0.19\linewidth]{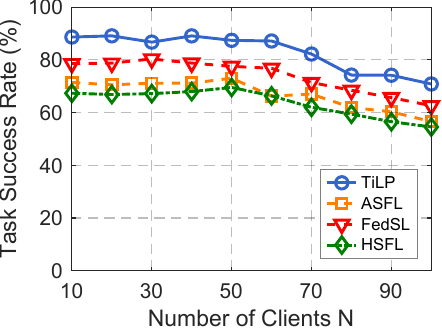}\label{fig:scale-success}}\hfil
    \subfloat[Latency]{\includegraphics[width=0.19\linewidth]{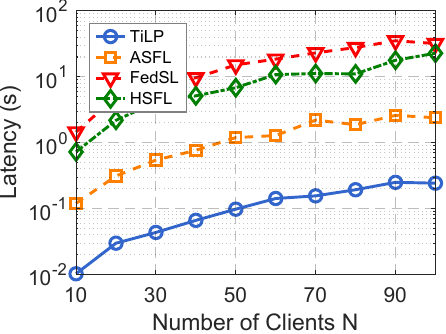}\label{fig:scale-latency}}\hfil
    \subfloat[Cumulative energy]{\includegraphics[width=0.19\linewidth]{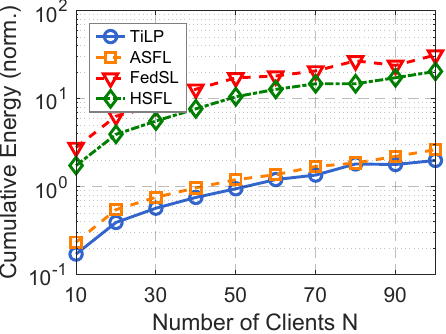}\label{fig:scale-energy}}\hfil
    \subfloat[Uplink SNR]{\includegraphics[width=0.19\linewidth]{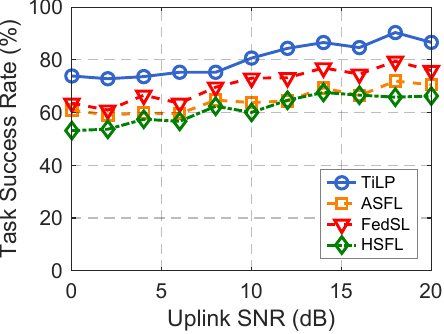}\label{fig:rob-snr}}\hfil
    \subfloat[Payload mass shift]{\includegraphics[width=0.19\linewidth]{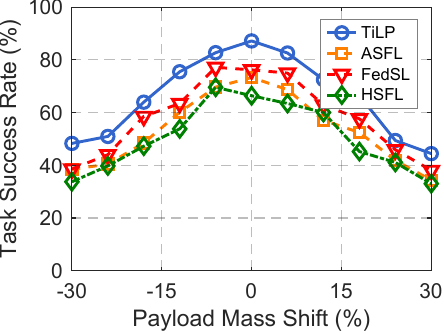}\label{fig:rob-mass}}
    \caption{Scalability and robustness of TiLP against ASFL, FedSL, and HSFL. Panels (a) to (c) vary the number of terminals from $N=10$ to $N=100$. Panels (d) and (e) test robustness to uplink SNR variation and payload-mass distribution shift.}
    \label{fig:scale-rob}
\end{figure*}

Fig.~\ref{fig:crosslayer-ablation} studies two important split-learning controls: activation retention and split-layer selection. The top row varies the retained activation fraction $1-q_n[t]$. A larger value preserves more activation information, which can improve task success. However, it also increases latency and uplink volume because more bits must be transmitted. TiLP remains better than the baselines across the retention sweep, which shows that its advantage does not come from a single hand-tuned compression level.

The bottom row varies the split-layer strategy. Moving the split layer changes both the client-side computation workload and the activation size. Therefore, the best layer for latency may not be the best layer for communication volume or task success. TiLP is less sensitive to this choice because it jointly adjusts the split layer, compression level, bandwidth, power, and scheduling decisions.

\subsection{Scalability and Robustness}
\label{sec:exp-scale}

Fig.~\ref{fig:scale-rob} evaluates scalability and robustness. Fig.~\ref{fig:scale-success} to Fig.~\ref{fig:scale-energy} increase the number of terminals. As $N$ grows, the system faces stronger terminal heterogeneity and a larger straggler effect. This usually increases latency and energy while making training less stable. TiLP maintains a clear advantage at large $N$ because it can schedule useful terminals and avoid repeatedly selecting terminals with poor wireless or learning conditions.

Fig.~\ref{fig:rob-snr} and Fig.~\ref{fig:rob-mass} test robustness to wireless and task changes. Under low uplink SNR, all methods suffer because activations and gradients become more expensive to transmit. TiLP degrades more gracefully because it can jointly adjust power, bandwidth, compression, and scheduling. Under payload-mass shift, the performance drop comes from task distribution mismatch rather than wireless impairment. This result shows why calibrating the task sub-twin is important.
\begin{table*}[t]
\centering
\setlength{\tabcolsep}{6pt}
\renewcommand\arraystretch{1.15}
\caption{Ablation studies of TiLP. ``w/o Network'', ``w/o Training'', and ``w/o Task'' freeze calibration Loop 1, Loop 2, and Loop 3 respectively while keeping all sub-twins active; ``w/o calibration'' freezes all three loops. ``w/o CEM'' bypasses CEM and outputs the actor's mean directly. ``w/o Task-driven'' replaces the task-success reward with the negative training-loss decrement, leaving twin and planner intact. ``--'' denotes an unreached threshold.}
\label{tab:ablation-studies}
\resizebox{0.98\linewidth}{!}{
\begin{tabular}{l|c|c|c|c|c|c}
\Xhline{1.2pt}
\textbf{Variant} & \textbf{Success (\%)} & \textbf{RTA (60/70/80)} & \textbf{Latency (s)} & \textbf{Energy (norm.)} & \textbf{Volume (GB)} & \textbf{Violation} \\
\Xhline{1.2pt}
Full                       & 87.5\std{2.04} & 29 / 68 / 144   & 0.10\std{0.014} & 1.00            & 0.42\std{0.013} & 3.52\std{0.024} \\
w/o Network                & 79.8\std{2.87} & 47 / 118 / --   & 0.16\std{0.018} & 1.21\std{0.037} & 0.58\std{0.021} & 5.84\std{0.071} \\
w/o Training               & 74.6\std{3.41} & 73 / 267 / --   & 0.21\std{0.024} & 1.38\std{0.052} & 0.71\std{0.028} & 7.92\std{0.094} \\
w/o Task                   & 80.7\std{2.63} & 42 / 109 / 348  & 0.15\std{0.014} & 1.17\std{0.029} & 0.55\std{0.018} & 5.41\std{0.063} \\
w/o Actor warm-start       & 82.1\std{2.18} & 38 / 94 / 284   & 0.13\std{0.012} & 1.11\std{0.026} & 0.51\std{0.016} & 4.76\std{0.047} \\
w/o Critic terminal        & 83.9\std{1.93} & 34 / 82 / 231   & 0.12\std{0.011} & 1.07\std{0.022} & 0.48\std{0.014} & 4.21\std{0.038} \\
w/o CEM                    & 78.4\std{3.06} & 54 / 154 / --   & 0.18\std{0.021} & 1.26\std{0.041} & 0.62\std{0.024} & 6.37\std{0.078} \\
w/o Task-driven            & 71.2\std{3.78} & 91 / 378 / --   & 0.24\std{0.029} & 1.47\std{0.063} & 0.79\std{0.034} & 8.94\std{0.116} \\
w/o calibration            & 64.3\std{4.52} & 138 / -- / --   & 0.34\std{0.041} & 1.78\std{0.087} & 1.03\std{0.048} & 12.67\std{0.183} \\
\Xhline{1.2pt}
\end{tabular}}
\end{table*}

Table~\ref{tab:ablation-studies} reports the ablation results. Among the three calibration loops, the training loop is the most important because TiLP depends directly on accurate loss-decrease estimates when evaluating candidate actions. The network loop is the second most important; stale channel estimates lead to inefficient bandwidth and power allocation, which increases energy consumption and constraint violation. The task loop has a milder effect because task-level performance changes more slowly than wireless conditions or training dynamics. Removing all three loops causes the largest degradation, indicating that the calibration loops are complementary rather than redundant. For the planner design, CEM provides the largest gain, while the actor warm start and critic terminal value offer additional but smaller improvements. Finally, replacing the task-success reward with the loss-decrease reward leads to a large performance drop. This observation confirms that training loss is not a reliable proxy for task success and supports the need for task-success-oriented planning.

\section{Conclusion}
\label{sec:conclusion}

This paper investigated task-success-oriented resource allocation for federated split learning at the wireless edge. Unlike conventional formulations that optimize training loss or communication cost alone, the proposed formulation directly accounts for the downstream task success rate while jointly controlling bandwidth, transmit power, split-layer selection, compression, and terminal scheduling. This setting is challenging because wireless dynamics, split-learning updates, and task-level performance evolve at different time scales and interact through a mixed continuous-discrete action space. 

To address this challenge, we proposed \method{}, a twin-in-the-loop planner that evaluates candidate actions through a cross-domain digital twin before applying them to the physical system. The digital twin integrates network, training, and task sub-twins, and each sub-twin is calibrated at the time scale of the process it models. On top of this twin, \method{} combines receding-horizon CEM planning with actor-critic learning, which enables efficient search over the coupled resource-allocation decisions.

Experimental results demonstrate that \method{} improves task success while reducing latency, energy consumption, uplink communication volume, and constraint violations compared with representative FSL baselines. The ablation study further shows that all three calibration loops are important: removing any one of them degrades planning quality, while removing all calibration leads to the largest performance loss. These results confirm that effective wireless FSL requires joint reasoning across communication, learning, and task layers.

Future work will extend this framework to larger-scale deployments with more heterogeneous devices, more dynamic task distributions, and broader classes of embodied-AI applications.



\begin{thebibliography}{10}
\providecommand{\url}[1]{#1}
\csname url@samestyle\endcsname
\providecommand{\newblock}{\relax}
\providecommand{\bibinfo}[2]{#2}
\providecommand{\BIBentrySTDinterwordspacing}{\spaceskip=0pt\relax}
\providecommand{\BIBentryALTinterwordstretchfactor}{4}
\providecommand{\BIBentryALTinterwordspacing}{\spaceskip=\fontdimen2\font plus
\BIBentryALTinterwordstretchfactor\fontdimen3\font minus
  \fontdimen4\font\relax}
\providecommand{\BIBforeignlanguage}[2]{{%
\expandafter\ifx\csname l@#1\endcsname\relax
\typeout{** WARNING: IEEEtran.bst: No hyphenation pattern has been}%
\typeout{** loaded for the language `#1'. Using the pattern for}%
\typeout{** the default language instead.}%
\else
\language=\csname l@#1\endcsname
\fi
#2}}
\providecommand{\BIBdecl}{\relax}
\BIBdecl

\bibitem{Singh2017PMLR}
\BIBentryALTinterwordspacing
B.~McMahan, E.~Moore, D.~Ramage, S.~Hampson, and B.~A.~y. Arcas,
  ``{Communication-Efficient Learning of Deep Networks from Decentralized
  Data},'' in \emph{Proceedings of the 20th International Conference on
  Artificial Intelligence and Statistics}, ser. Proceedings of Machine Learning
  Research, A.~Singh and J.~Zhu, Eds., vol.~54.\hskip 1em plus 0.5em minus
  0.4em\relax PMLR, 20--22 Apr 2017, pp. 1273--1282. [Online]. Available:
  \url{https://proceedings.mlr.press/v54/mcmahan17a.html}
\BIBentrySTDinterwordspacing

\bibitem{Praneeth2018arXiv}
\BIBentryALTinterwordspacing
P.~Vepakomma, O.~Gupta, T.~Swedish, and R.~Raskar, ``Split learning for health:
  Distributed deep learning without sharing raw patient data,'' vol.
  abs/1812.00564, 2018. [Online]. Available:
  \url{http://arxiv.org/abs/1812.00564}
\BIBentrySTDinterwordspacing

\bibitem{Thapa2022AAAI}
C.~Thapa, P.~C. Mahawaga~Arachchige, S.~Camtepe, and L.~Sun, ``{SplitFed: When
  Federated Learning Meets Split Learning},'' \emph{Proceedings of the AAAI
  Conference on Artificial Intelligence}, vol.~36, no.~8, pp. 8485--8493, 2022.

\bibitem{Wu2026MNET}
B.~Wu, J.~Huang, and Q.~Duan, ``{Real-Time Intelligent Healthcare Enabled by
  Federated Digital Twins With AoI Optimization},'' \emph{IEEE Network},
  vol.~40, no.~2, pp. 184--191, 2025.

\bibitem{Wu2026COMST}
B.~Wu, J.~Huang, and S.~Yu, ``{``X of Information'' Continuum: A Survey on
  AI-Driven Multi-Dimensional Metrics for Next-Generation Networked Systems},''
  \emph{IEEE Communications Surveys \& Tutorials}, vol.~28, pp. 5307--5344,
  2026.

\bibitem{DingICNC2025}
Z.~Ding, J.~Huang, and J.~Qi, ``{Learning to Defend: A Multi-Agent
  Reinforcement Learning Framework for Stackelberg Security Game in Mobile Edge
  Computing},'' in \emph{2026 International Conference on Computing, Networking
  and Communications (ICNC)}, 2026, pp. 769--774.

\bibitem{Wu2025ToN}
B.~Wu, J.~Huang, Q.~Duan, L.~Dong, and Z.~Cai, ``{Enhancing Vehicular
  Platooning With Wireless Federated Learning: A Resource-Aware Control
  Framework},'' \emph{IEEE/ACM Transactions on Networking}, pp. 1--1, 2025.

\bibitem{Wu2026TNSE}
B.~Wu, Z.~Ding, and J.~Huang, ``{A Review of Continual Learning in Edge AI},''
  \emph{IEEE Transactions on Network Science and Engineering}, vol.~13, pp.
  6571--6588, 2026.

\bibitem{Xu2024TWC}
C.~Xu, J.~Li, Y.~Liu, Y.~Ling, and M.~Wen, ``{Accelerating Split Federated
  Learning Over Wireless Communication Networks},'' \emph{IEEE Transactions on
  Wireless Communications}, vol.~23, no.~6, pp. 5587--5599, 2024.

\bibitem{Lin2025TMC_a}
Z.~Lin, W.~Wei, Z.~Chen, C.-T. Lam, X.~Chen, Y.~Gao, and J.~Luo,
  ``{Hierarchical Split Federated Learning: Convergence Analysis and System
  Optimization},'' \emph{IEEE Transactions on Mobile Computing}, vol.~24,
  no.~10, pp. 9352--9367, 2025.

\bibitem{Hou2025JSAC}
X.~Hou, J.~Wang, F.~Guan, J.~Du, and C.~Jiang, ``{Energy-Efficient Federated
  Learning for Edge Real-Time Vision via Joint Data, Computation, and
  Communication Design},'' \emph{IEEE Journal on Selected Areas in
  Communications}, vol.~43, no.~12, pp. 4000--4014, 2025.

\bibitem{Wang2025TWC}
Z.~Wang, Y.~Zhou, Y.~Shi, and K.~B. Letaief, ``{Federated Fine-Tuning for
  Pre-Trained Foundation Models Over Wireless Networks},'' \emph{IEEE
  Transactions on Wireless Communications}, vol.~24, no.~4, pp. 3450--3464,
  2025.

\bibitem{Wang2026TWC}
T.~Wang, Y.~Zhou, Y.~Shi, N.~Cheng, and H.~Zhou, ``{Zeroth-Order Federated
  Fine-Tuning for Large AI Models in Resource-Constrained Wireless Networks},''
  \emph{IEEE Transactions on Wireless Communications}, vol.~25, pp.
  12\,407--12\,421, 2026.

\bibitem{Wu2023MPE}
B.~Wu and W.~Wu, ``{Model-Free Cooperative Optimal Output Regulation for Linear
  Discrete-Time Multi-Agent Systems Using Reinforcement Learning},''
  \emph{Mathematical Problems in Engineering}, vol. 2023, no.~1, p. 6350647,
  2023.

\bibitem{Wu2025RACS}
B.~Wu, Z.~Ding, L.~Ostigaard, and J.~Huang, ``{Reinforcement Learning-Based
  Energy-Aware Coverage Path Planning for Precision Agriculture},'' in
  \emph{2025 ACM Research on Adaptive and Convergent Systems (RACS)}.\hskip 1em
  plus 0.5em minus 0.4em\relax ACM, 2025, pp. 1--8.

\bibitem{Ding2025IPCCC}
Z.~Ding, J.~Huang, Q.~Duan, C.~Zhang, Y.~Zhao, and S.~Gu, ``{A Dual-Level
  Game-Theoretic Approach for Collaborative Learning in UAV-Assisted
  Heterogeneous Vehicle Networks},'' in \emph{2025 IEEE International
  Performance, Computing, and Communications Conference (IPCCC)}, 2025, pp.
  1--8.

\bibitem{Pudasaini2026HPSR}
U.~Pudasaini, Z.~Ding, and J.~Huang, ``{Securing Smart Agriculture with
  Communication-Efficient Federated Unlearning},'' in \emph{2026 IEEE
  International Conference on High Performance Switching and Routing
  (HPSR)}.\hskip 1em plus 0.5em minus 0.4em\relax IEEE, 2026, pp. 1--8.

\bibitem{Khan2025TWC}
N.~Khan, A.~Abdallah, A.~\c{C}elik, A.~M. Eltawil, and S.~\c{C}\"oleri,
  ``{Digital Twin-Assisted Explainable AI for Robust Beam Prediction in mmWave
  MIMO Systems},'' \emph{IEEE Transactions on Wireless Communications},
  vol.~25, pp. 2435--2451, 2025.

\bibitem{Okegbile2025TMC}
S.~D. Okegbile, H.~Gao, and J.~Cai, ``{A Novel Secure Split Federated Semantic
  Learning Framework and its Optimization for Digital Twin Network
  Evolution},'' \emph{IEEE Transactions on Mobile Computing}, vol.~25, no.~1,
  pp. 1302--1319, 2025.

\bibitem{Zhou2025TMC}
L.~Zhou, S.~Leng, Y.~Liu, Z.~Xiong, and T.~Q.~S. Quek, ``{Digital Twins for
  Low-Altitude UAV Networks: Cooperation and Learning},'' \emph{IEEE
  Transactions on Mobile Computing}, vol.~25, no.~4, pp. 4839--4856, 2025.

\bibitem{Ho2025JSAC}
T.~M. Ho, K.~K. Nguyen, and M.~Cheriet, ``{AI-Powered Digital Twins for Robotic
  Control in 5G-Enabled Industrial Automation},'' \emph{IEEE Journal on
  Selected Areas in Communications}, vol.~43, no.~10, pp. 3347--3361, 2025.

\bibitem{Wu2026ICDCS}
B.~Wu, J.~Huang, and Y.~Zhao, ``{From Alpha to Omega: Lifecycle-Aware
  Forgetting Defense in Federated Continual Learning for Planetary
  Exploration},'' in \emph{Proceedings of the IEEE International Conference on
  Distributed Computing Systems (ICDCS)}, 2026.

\bibitem{Ding2026ICDCS}
Z.~Ding, B.~Wu, and J.~Huang, ``{SCALE: Sensitivity-Aware Federated Unlearning
  with Information Freshness Optimization for Mobile Edge Computing},'' in
  \emph{Proceedings of the IEEE International Conference on Distributed
  Computing Systems (ICDCS)}, 2026.

\bibitem{Hafner2019ICML}
D.~Hafner, T.~Lillicrap, I.~Fischer, R.~Villegas, D.~Ha, H.~Lee, and
  J.~Davidson, ``{Learning Latent Dynamics for Planning from Pixels},'' in
  \emph{Proceedings of the International Conference on Machine Learning}, 2019,
  pp. 2555--2565.

\bibitem{Hafner2025NATURE}
D.~Hafner, J.~Pa\v{s}ukonis, J.~Ba, and T.~Lillicrap, ``{Mastering Diverse
  Control Tasks through World Models},'' \emph{Nature}, vol. 640, pp. 647--653,
  2025.

\bibitem{Chua2018NeurIPS}
K.~Chua, R.~Calandra, R.~McAllister, and S.~Levine, ``{Deep Reinforcement
  Learning in a Handful of Trials using Probabilistic Dynamics Models},'' in
  \emph{Advances in Neural Information Processing Systems}, 2018, pp.
  4754--4765.

\bibitem{Janner2019NeurIPS}
M.~Janner, J.~Fu, M.~Zhang, and S.~Levine, ``{When to Trust Your Model:
  Model-Based Policy Optimization},'' in \emph{Advances in Neural Information
  Processing Systems}, 2019, pp. 12\,519--12\,530.

\bibitem{Huang2025TMC}
J.~Huang, B.~Wu, Q.~Duan, L.~Dong, and S.~Yu, ``{A Fast UAV Trajectory Planning
  Framework in RIS-Assisted Communication Systems With Accelerated Learning via
  Multithreading and Federating},'' \emph{IEEE Transactions on Mobile
  Computing}, pp. 1--16, 2025.

\bibitem{Wu2026ARXIV}
B.~Wu, Z.~Ding, and J.~Huang, ``{RELIEF: Turning Missing Modalities into
  Training Acceleration for Federated Learning on Heterogeneous IoT Edge},''
  arXiv preprint arXiv:2604.04243, 2026.

\bibitem{Fang2025ARXIV}
Z.~Fang, Y.~Guo, J.~Wang, Y.~Zhang, H.~An, Y.~Wang, and Y.~Fang, ``{Shared
  Spatial Memory Through Predictive Coding},'' arXiv preprint arXiv:2511.04235,
  2025.

\bibitem{Fang2026GLOBECOM}
Z.~Fang, Z.~Liu, J.~Wang, S.~Hu, Y.~Guo, Y.~Deng, and Y.~Fang, ``Task-oriented
  communications for visual navigation with edge-aerial collaboration in low
  altitude economy,'' in \emph{Proc. IEEE Global Communications Conference
  (GLOBECOM)}, 2026.

\bibitem{Mu2025TMC}
Y.~Mu and C.~Shen, ``{Federated Split Learning With Improved Communication and
  Storage Efficiency},'' \emph{IEEE Transactions on Mobile Computing}, vol.~25,
  no.~1, pp. 272--283, 2025.

\bibitem{Ao2026TWC}
H.~Ao, H.~Tian, W.~Ni, J.~Zhang, and D.~Niyato, ``{Federated Split Learning via
  Low-Rank Approximation: A Communication-Efficient Approach},'' \emph{IEEE
  Transactions on Wireless Communications}, vol.~25, pp. 11\,253--11\,269,
  2026.

\bibitem{Qiao2024JSAC}
L.~Qiao, Z.~Gao, M.~Boloursaz~Mashhadi, and D.~G\"und\"uz, ``{Massive Digital
  Over-the-Air Computation for Communication-Efficient Federated Edge
  Learning},'' \emph{IEEE Journal on Selected Areas in Communications},
  vol.~42, no.~11, pp. 3078--3094, 2024.

\bibitem{Wu2023ACCESS}
B.~Wu, Z.~Cai, W.~Wu, and X.~Yin, ``{AoI-Aware Resource Management for Smart
  Health via Deep Reinforcement Learning},'' \emph{IEEE Access}, 2023.

\bibitem{Pan2023SCIS}
D.~Pan, B.-N. Wu, Y.-L. Sun, and Y.-P. Xu, ``{A Fault-Tolerant and
  Energy-Efficient Design of a Network Switch Based on a Quantum-Based
  Nano-Communication Technique},'' \emph{Sustainable Computing: Informatics and
  Systems}, vol.~37, p. 100827, 2023.

\bibitem{Lin2025TMC_b}
Z.~Lin, G.~Qu, W.~Wei, X.~Chen, and K.~K. Leung, ``{AdaptSFL: Adaptive Split
  Federated Learning in Resource-Constrained Edge Networks},'' \emph{IEEE
  Transactions on Networking}, vol.~33, no.~6, pp. 2993--3008, 2025.

\bibitem{Ao2025TCE}
H.~Ao, H.~Tian, and W.~Ni, ``{Federated Split Learning for Edge Intelligence in
  Resource-Constrained Wireless Networks},'' \emph{IEEE Transactions on
  Consumer Electronics}, vol.~71, no.~2, pp. 4451--4463, 2025.

\bibitem{Liang2026TWC}
Y.~Liang, Q.~Chen, R.~Li, G.~Zhu, M.~Kaleem~Awan, and H.~Jiang,
  ``{Communication-and-Computation Efficient Split Federated Learning in
  Wireless Networks: Gradient Aggregation and Resource Management},''
  \emph{IEEE Transactions on Wireless Communications}, vol.~25, pp. 1981--1995,
  2026.

\bibitem{You2026TMC}
J.~You, J.~Yan, Z.~Li, and L.~Yang, ``{Adaptive Bayesian Optimization for
  Online Bandit Model Partitioning and Resource Allocation in Split Federated
  Learning},'' \emph{IEEE Transactions on Mobile Computing}, 2026.

\bibitem{Xie2025TWC}
C.~Xie, Z.~Chen, W.~Yi, H.~Shin, and A.~Nallanathan, ``{Tackling Class
  Imbalance and Client Heterogeneity for Split Federated Learning in Wireless
  Networks},'' \emph{IEEE Transactions on Wireless Communications}, vol.~24,
  no.~6, pp. 4920--4936, 2025.

\bibitem{Wang2024TON}
S.~Wang, R.~Morabito, S.~Hosseinalipour, M.~Chiang, and C.~G. Brinton,
  ``{Device Sampling and Resource Optimization for Federated Learning in
  Cooperative Edge Networks},'' \emph{IEEE/ACM Transactions on Networking},
  vol.~32, no.~5, pp. 4365--4381, 2024.

\bibitem{Wen2025TNSE}
Y.~Wen, G.~Zhang, K.~Wang, and K.~Yang, ``{Training Latency Minimization for
  Model-Splitting Allowed Federated Edge Learning},'' \emph{IEEE Transactions
  on Network Science and Engineering}, vol.~12, no.~3, pp. 2081--2092, 2025.

\bibitem{Xing2026ACR}
\BIBentryALTinterwordspacing
C.-C. Xing, Z.~Ding, and J.~Huang, ``{A Stochastic Geometry-Based Analysis of
  SWIPT-Assisted Underlaid Device-to-Device Energy Harvesting},'' \emph{SIGAPP
  Appl. Comput. Rev.}, vol.~25, no.~4, p. 18–34, Jan. 2026. [Online].
  Available: \url{https://doi.org/10.1145/3787594.3787596}
\BIBentrySTDinterwordspacing

\bibitem{Cao2025TWC}
Y.~Cao, J.~Wang, X.~Shi, and W.~Ni, ``{Lightweight and Self-Evolving Channel
  Twinning: An Ensemble DMD-Assisted Approach},'' \emph{IEEE Transactions on
  Wireless Communications}, vol.~24, no.~10, pp. 8072--8085, 2025.

\bibitem{Zhang2024JSAC}
Z.~Zhang, Y.~Liu, Z.~Peng, M.~Chen, and D.~Xu, ``{Digital Twin-Assisted
  Data-Driven Optimization for Reliable Edge Caching in Wireless Networks},''
  \emph{IEEE Journal on Selected Areas in Communications}, vol.~42, no.~11, pp.
  3306--3320, 2024.

\bibitem{Wu2026ARXIV1}
B.~Wu and J.~Huang, ``{Lifecycle-Aware Federated Continual Learning in Mobile
  Autonomous Systems},'' arXiv preprint arXiv:2604.20745, 2026.

\bibitem{Fang2025TON}
Z.~Fang, S.~Hu, J.~Wang, Y.~Deng, X.~Chen, and Y.~Fang, ``{Prioritized
  Information Bottleneck Theoretic Framework With Distributed Online Learning
  for Edge Video Analytics},'' \emph{IEEE Transactions on Networking}, pp.
  1--17, 2025.

\bibitem{Fang2025JSAC}
Z.~Fang, J.~Wang, Y.~Ma, Y.~Tao, Y.~Deng, X.~Chen, and Y.~Fang, ``{R-ACP:
  Real-Time Adaptive Collaborative Perception Leveraging Robust Task-Oriented
  Communications},'' \emph{IEEE Journal on Selected Areas in Communications},
  2025.

\bibitem{Jin2026TNSE}
D.~Jin, Y.~Xiao, Y.~Li, and G.~Shi, ``{Personalized Federated Learning for
  Generative AI Empowered Digital Twin Networks},'' \emph{IEEE Transactions on
  Network Science and Engineering}, vol.~13, pp. 6174--6192, 2026.

\bibitem{Chen2025TMC}
X.~Chen, J.~Cao, R.~Cao, Y.~Sahni, and M.~Zhang, ``{Decentralized Task
  Offloading in Collaborative Edge Computing: A Digital Twin Assisted
  Multi-Agent Reinforcement Learning Approach},'' \emph{IEEE Transactions on
  Mobile Computing}, vol.~25, no.~4, pp. 4776--4790, 2025.

\bibitem{Chen2024TMC}
X.~Chen, J.~Cao, Y.~Sahni, M.~Zhang, and Z.~Liang, ``{Mobility-Aware Dependent
  Task Offloading in Edge Computing: A Digital Twin-Assisted Reinforcement
  Learning Approach},'' \emph{IEEE Transactions on Mobile Computing}, vol.~24,
  no.~4, pp. 2979--2994, 2024.

\bibitem{Rubinstein1999}
R.~Y. Rubinstein, ``{The Cross-Entropy Method for Combinatorial and Continuous
  Optimization},'' \emph{Methodology and Computing in Applied Probability},
  vol.~1, no.~2, pp. 127--190, 1999.

\bibitem{Haarnoja2018ICML}
T.~Haarnoja, A.~Zhou, P.~Abbeel, and S.~Levine, ``{Soft Actor-Critic:
  Off-Policy Maximum Entropy Deep Reinforcement Learning with a Stochastic
  Actor},'' in \emph{Proceedings of the International Conference on Machine
  Learning}, 2018, pp. 1861--1870.

\bibitem{Tong2025TMC}
H.~Tong, M.~Chen, J.~Zhao, Y.~Hu, and Z.~Yang, ``{Continual Reinforcement
  Learning for Digital Twin Synchronization Optimization},'' \emph{IEEE
  Transactions on Mobile Computing}, vol.~24, no.~8, pp. 6843--6857, 2025.

\bibitem{Liu2024TMC}
Z.~Liu, H.~Du, J.~Lin, Z.~Gao, and L.~Huang, ``{DNN Partitioning, Task
  Offloading, and Resource Allocation in Dynamic Vehicular Networks: A
  Lyapunov-Guided Diffusion-Based Reinforcement Learning Approach},''
  \emph{IEEE Transactions on Mobile Computing}, vol.~24, no.~3, pp. 1945--1962,
  2024.

\bibitem{Wang2025MCOM}
J.~Wang, J.~Zhang, Y.~Zhang, Y.~Sun, G.~Nie, L.~Shi, P.~Zhang, and G.~Liu,
  ``{Radio Environment Knowledge Pool for 6G Digital Twin Channel},''
  \emph{IEEE Communications Magazine}, vol.~63, no.~5, pp. 158--164, 2025.

\bibitem{Liu2023NeurIPS}
B.~Liu, Y.~Zhu, C.~Gao, Y.~Feng, Q.~Liu, Y.~Zhu, and P.~Stone, ``{LIBERO:
  Benchmarking Knowledge Transfer for Lifelong Robot Learning},'' in
  \emph{Advances in Neural Information Processing Systems}, 2023.

\bibitem{Yang2021TWC}
Z.~Yang, M.~Chen, W.~Saad, C.~S. Hong, and M.~Shikh-Bahaei, ``{Energy Efficient
  Federated Learning Over Wireless Communication Networks},'' \emph{IEEE
  Transactions on Wireless Communications}, vol.~20, no.~3, pp. 1935--1949,
  2021.

\bibitem{Jennifer2018PMLR}
\BIBentryALTinterwordspacing
T.~Haarnoja, A.~Zhou, P.~Abbeel, and S.~Levine, ``{Soft Actor-Critic:
  Off-Policy Maximum Entropy Deep Reinforcement Learning with a Stochastic
  Actor},'' in \emph{Proceedings of the 35th International Conference on
  Machine Learning}, ser. Proceedings of Machine Learning Research, J.~Dy and
  A.~Krause, Eds., vol.~80.\hskip 1em plus 0.5em minus 0.4em\relax PMLR, 10--15
  Jul 2018, pp. 1861--1870. [Online]. Available:
  \url{https://proceedings.mlr.press/v80/haarnoja18b.html}
\BIBentrySTDinterwordspacing

\bibitem{Eric2017ICLR}
\BIBentryALTinterwordspacing
E.~Jang, S.~Gu, and B.~Poole, ``{Categorical Reparameterization with
  Gumbel-Softmax},'' in \emph{5th International Conference on Learning
  Representations, {ICLR} 2017, Toulon, France, April 24-26, 2017, Conference
  Track Proceedings}.\hskip 1em plus 0.5em minus 0.4em\relax OpenReview.net,
  2017. [Online]. Available: \url{https://openreview.net/forum?id=rkE3y85ee}
\BIBentrySTDinterwordspacing

\end{thebibliography}

\end{document}